\documentclass[12pt, preprint]{aastex}

\begin{document}

\input epsf.tex    

\input psfig.sty

\include{epsfig}



\def\etal{et al.\ } 
\def\msun{$M_\odot$\ } 
\def\msunp{$M_\odot$} 
\def\yr{yr$^{-1}$\ }
\def\lsun{L$_\odot$\ } 
\def\lsunp{L$_\odot$} 
\def\kms {km~s$^{-1}$\ } 
\def\kmsp{km~s$^{-1}$} 
\def\jhk{\hbox{$J\!H\!K$}} 
\def\jhkl{\hbox{$J\!HK\!L$}}
\def\hk{\hbox{$H\!-\!K$}} 
\def\jh{\hbox{$J\!-\!H$}} 
\def\kl{\hbox{$K\!-\!L$}}
\def\co{$^{12}$CO\ } 
\def\h2{H$_2$\ } 
\def\cmv{cm$^{-3}$} 
\def\pc3{pc$^{-3}$} 
\def\um{$\mu$m} 
\def\h {$^h$} 
\def\m {$^m$} 
\def\s {$^s$}
\def\deg{$^{\circ}$} 
\def\degs{$^{\circ}$} 
\def\am{${'}$} 

\slugcomment{To appear in Annual Reviews of Astronomy \& Astrophysics}

\title{Embedded Clusters in Molecular Clouds}

\markboth{Lada \& Lada}{Embedded Clusters}

\author{Charles J.  Lada} 
\affil{Harvard-Smithsonian Center for Astrophysics, 60
Garden Street, Cambridge, Massachusetts 02138; email:  clada@cfa.harvard.edu} 
\author{Elizabeth A.  Lada} 
\affil{Department of Astronomy, University of Florida, Gainesville,
Florida 32611; email:  lada@astro.ufl.edu}


\begin{abstract}

Stellar clusters are born embedded within giant molecular clouds (GMCs) and during their
formation and early evolution are often only visible at infrared wavelengths, being heavily
obscured by dust.  Over the last 15 years advances in infrared detection capabilities have
enabled the first systematic studies of embedded clusters in galactic molecular clouds.  In this
article we review the current state of empirical knowledge concerning these extremely young
protocluster systems.  From a survey of the literature we compile the first extensive catalog of
galactic embedded cluster properties.  We use the catalog to construct the mass function and
estimate the birthrate for embedded clusters within $\sim$ 2 Kpc of the Sun.  We find that the
embedded cluster birthrate exceeds that of visible open clusters by an order of magnitude or more
indicating a high infant mortality rate for protocluster systems.  Less than 4-7\% of embedded
clusters survive emergence from molecular clouds to become bound clusters of Pleiades age.  The
vast majority (90\%) of stars that form in embedded clusters form in rich clusters of 100 or more
members with masses in excess of 50\msunp.  Moreover, observations of nearby cloud complexes
indicate that embedded clusters account for a significant (70-90\%) fraction of all stars formed
in GMCs.  We review the role of embedded clusters in investigating the nature of the IMF which,
in one nearby example, has been measured over the entire range of stellar and substellar mass,
from OB stars to subsellar objects near the deuterium burning limit.  We also review the role
embedded clusters play in the investigation of circumstellar disk evolution and the important
constraints they provide for understanding the origin of planetary systems.  Finally, we discuss
current ideas concerning the origin and dynamical evolution of embedded clusters and the
implications for the formation of bound open clusters.

\end{abstract}

\tableofcontents

\maketitle

\section{INTRODUCTION}

Stellar clusters have been long recognized as important laboratories for astrophysical
research.  Their study has played an important role in developing an understanding of
the universe.  For example, clusters contain statistically significant samples of stars
spanning a wide range of stellar mass within a relatively small volume of space.  Since
stars in such groups share the common heritage of being formed more or less
simultaneously from the same progenitor molecular cloud, observations of cluster
color-magnitude (CM) diagrams can be, and indeed, have been used to provide classical
tests of stellar evolution theory.  Moreover, clusters offer the smallest physical scale
over which a meaningful determination of the stellar Initial Mass Function (IMF) can be
made.  Because a cluster is held together by the mutual gravitational attraction of its
individual members, its evolution is determined by Newton's laws of motion and gravity.
In many body systems these interactions are inherently complex and thus clusters are
also important testbeds for studies of stellar dynamics.  The spatial distribution of
clusters has also played a vital role in our understanding of galactic structure.  The
distribution of globular clusters, for example, was critical for determining the
location of the galactic center, establishing the existence of a galactic halo and
setting the overall scale of the Galaxy.  Young open clusters have provided an important
tracer of recent star formation in galaxies and of spiral structure in galactic disks.
Such clusters are also of interest for understanding the origin of the solar system,
since the presence of rare short-lived radio nuclides in meteoritic samples has long
suggested that the Sun itself was formed in near proximity to a massive star, and thus
most likely within in relatively rich cluster.

Little is known or understood about the origin of clusters.  Globular clusters in the
Galaxy were formed billions of years ago.  Because they are not being formed in the Milky
Way in the present epoch of galactic history, direct empirical study of their formation
process is not possible (except perhaps in certain extragalactic systems and at
cosmological distances).  On the other hand, open clusters appear to be continuously
forming in the galactic disk and, in principle, direct study of the physical processes
leading to their formation is possible.  However, such studies have been seriously hampered
by the fact that galactic clusters form in giant molecular clouds (GMCs) and during their
formation and earliest stages of evolution are completely embedded in molecular gas and
dust, and thus obscured from view.  Given the constraints imposed by traditional techniques
of optical astronomy, direct observation and study of young embedded clusters had been
extremely difficult, if not impossible.  However, during the last two decades the
development of infrared astronomy and, more recently, infrared array detectors, has
dramatically improved this situation.  Figure~\ref{rcw38} shows optical and infrared images
of the southern embedded cluster RCW 38 and amply illustrates the power of infrared imaging
for detecting such heavily obscured young clusters.

\begin{figure} 
\vskip 6.0in

\epsfxsize=3.8in 
\caption{Optical (top) and Infrared (bottom) images of
the RCW 38 region obtained with the ESO VLT.  The infrared observations reveal a rich
embedded cluster otherwise invisible at optical wavelengths. Figure courtesy of J. Alves.}  
\label{rcw38}
\end{figure}

The deployment of infrared imaging cameras and spectrometers on optical and infrared
optimized telescopes has provided astronomers the ability to survey and systematically
study embedded clusters within molecular clouds.  Almost immediately such studies
indicated that rich embedded clusters were surprisingly numerous and that a significant
fraction, if not the vast majority, of all stars may form in such systems.
Consequently, it is now recognized that embedded clusters may be basic units of star
formation and their study can directly address a number of fundamental astrophysical
problems.  These include the issues of cluster formation and early evolution as well as
the more general problems of the origin and early evolution of stars and planetary
systems.  Because most stars in the galactic disk may originate in embedded clusters,
these systems must play a critical role in understanding the origins of some of the most
fundamental properties of the galactic stellar population, such as the form and
universality of the stellar IMF and the frequencies of stellar and planetary companions.

The purpose of this review is to summarize the current status of observational knowledge
concerning young embedded clusters in the Galaxy.  We will consider both embedded and
partially embedded clusters.  The embedded phase of cluster evolution appears to
last between 2-3 Myrs and clusters with ages greater than 5 Myrs are rarely associated with
molecular gas (Leisawitz, Bash \& Thaddeus 1989) therefore, this review deals with clusters
whose ages are typically between 0.5-3 million years.  Particular emphasis will be placed
on embedded clusters within $\sim$ 2 Kpc of the sun since this presents a sample which is
most statistically complete and for which the most detailed observational data are
available.  Previous reviews of embedded clusters, some with slightly different emphasis,
can be found in various conference proceedings (e.g., Lada \& Lada 1991; Zinnecker,
McCaughrean \& Wilking, 1993; Lada 1998; Clarke, Bonnell \& Hillenbrand 2000, Elmegreen et
al.  2000, Lada et al.  2002).

\section{EMBEDDED CLUSTERS: BASIC OBSERVATIONAL DATA}

\subsection{Definitions \& Terminology}

For the purposes of this review we consider clusters to be groups of stars which are
physically related and whose observed {\it stellar} mass volume density would be
sufficiently large, if in a state of virial equilibrium, to render the group stable against
tidal disruption by the galaxy (i.e., $\rho_* \geq 0.1$ \msun pc$^{-3}$; Bok 1934), and by
passing interstellar clouds (i.e., $\rho_* \geq 1.0 $ \msun pc$^{-3}$; Spitzer 1958).
Furthermore we adopt the additional criterion (e.g., Adams \& Myers 2001) that the cluster
consist of enough members to insure that its evaporation time (i.e., the time it takes for
internal stellar encounters to eject all its members) be greater than 10$^8$ yrs, the
typical lifetime of open clusters in the field.  The evaporation time, $\tau_{ev}$, for a
stellar system in virial equilibrium, is of order $\tau_{ev} \approx 10^2 \tau_{relax}$,
where the relaxation time is roughly $\tau_{relax} \approx {0.1N \over ln N} \tau_{cross}$
and $\tau_{cross}$ is the dynamical crossing time of the system and $N$ the number of stars
it contains (Binney \& Tremaine 1987).  The typical crossing time in open clusters is of
order 10$^6$ yrs, so if such a cluster is to survive disintegration by evaporation for
10$^8$ yrs, its relaxation time must be comparable or to or greater than its crossing time
or ${0.1N \over ln N} \approx 1$.  This condition is met when $N \approx 35$.  Therefore,
for this review, we define a stellar cluster as:  {\it a group of 35 or more physically
related stars whose stellar mass density exceeds 1.0 \msun pc$^{-3}$.}

With our definition we 
distinguish clusters from multiple star systems, such as small (N $<$ 6) hierarchical
multiples and binaries which are relatively stable systems and small multiple systems of
the Trapezium type, which are inherently unstable (Ambartsumian 1954; Allen \& Poveda
1974).  We also distinguish clusters from stellar associations, which we define as 
loose groups of physically related stars whose stellar space density is considerably
below the tidal stability limit of 1 \msun yr$^{-3}$ (Blaauw 1964).

Clusters, as defined above, can be classified into two environmental classes depending on
their association with interstellar matter.  {\it Exposed clusters} are clusters with
little or no interstellar matter within their boundaries.  Almost all clusters found in
standard open cluster catalogs (e.g., Lynga 1987) fall into this category.  {\it Embedded
clusters} are clusters which are fully or partially embedded in interstellar gas and dust.
They are frequently completely invisible at optical wavelengths and best detected in the
infrared.  These clusters are the youngest known stellar systems and can also be considered
protoclusters, since upon emergence from molecular clouds they will become exposed
clusters.  A similar classification can be applied to associations.

Our definition of a cluster includes stellar systems of two dynamical types or states.
{\it Bound clusters} are systems whose total energy (kinetic + potential) is negative.
When determining the total energy we include contributions from any interstellar material
contained within the boundaries of the cluster.  We define a classical open cluster as a
bound,
exposed cluster, such as the Pleiades, which can live to be at least 10$^8$ yrs in the
vicinity of the sun.  {\it Unbound clusters} are systems whose total energy is positive.
That is, unbound clusters are clusters of 35 or more stars whose space densities exceed 1
\msun pc$^{-3}$ but whose internal stellar motions are too large to be gravitationally
confined by the stellar and non-stellar material within the boundaries of the cluster.

\subsection{Identification \& Surveys}

Infrared surveys of molecular clouds are necessary to reveal embedded clusters, since
many if not all of their members will be heavily obscured. The initial identification of 
an embedded cluster is typically made by a survey at a single infrared wavelength
(e.g., 2.2$\mu$m or K-band). The existence of a cluster is established by an excess 
density of stars over the background. In general the ease of identifying a cluster 
depends sensitively on the richness of the cluster, the apparent brightness of its 
members, its angular size or compactness, its location in the Galactic plane and
the amount of obscuration in its direction. For example, it would be particularly 
difficult to recognize a spatially extended, poor cluster of faint stars located in a
direction where there is a high background of infrared sources, (e.g., l = 0.0, b = 0.0).

Identification of the individual members of a cluster is considerably more difficult than
establishing its existence.  In particular, for most clusters the source density of
intrinsically faint members is usually only comparable to or even significantly less than
that of background/foreground field objects.  In such circumstances cluster membership can
be determined only on a statistical basis, by comparison with star counts in nearby control
fields off the cluster.  However, determining whether or not a specific star in the region
is a cluster member or not is not generally possible from a star counting survey alone.  In
situations where field star contamination is non-negligible, other independent information
(e.g., proper motions, spectra, multi-wavelength photometry) is required to determine
membership of individual stars.

The first deeply embedded cluster identified in a molecular cloud was uncovered in
near-infrared surveys of the Ophiuchi dark cloud first made nearly thirty years ago using
single-channel infrared photometers (Grasdalen, Strom \& Strom 1974; Wilking \& Lada 1983).
However, it wasn't until the deployment of infrared imaging cameras in the late 1980s, that
large numbers of embedded clusters were identified and studied.  In a search of the
astronomical literature since 1988 we have found that well over a hundred such clusters
have been observed both nearby the sun (e.g., Eiroa \& Casali 1992) and at the distant
reaches of the galaxy (e.g., Santos et al.  2000).  To date embedded clusters have been
discovered using three basic observational approaches:  1) case studies of individual star
forming regions, such as for example, NGC 2282 (Horner, Lada \& Lada 1997), LKH$\alpha$ 101
(Barsony, Schombert \& Kis-Halas 1991), and NGC 281 (Megeath \& Wilson 1997), 2) systematic
surveys of various signposts of star formation, such as outflows (Hodapp 1994), luminous
IRAS sources (e.g., Carpenter et al.  1993), and Herbig AeBe stars (Testi, Natta \& Palla
1998), and 3) systematic surveys of individual molecular cloud complexes (e.g., Lada et al.
1991b; Carpenter Snell \& Schloerb 1995; Phelps \& Lada 1997; Carpenter, Heyer \& Snell
2000; Carpenter 2000).  To date most known embedded clusters have been found in surveys of
star formation signposts (2), in particular the Hodapp (1994) survey of outflows has had by
far the most prolific success rate.  In the near future, we expect surveys conducted using
the data generated by the all sky near-infrared surveys (i.e., DENIS and 2MASS) will likely
provide the the most systematic and complete inventory of the embedded cluster population
of the Galaxy.

\subsection{The Embedded Cluster Catalog}

We have compiled a catalog of embedded clusters within $\sim$ 2 Kpc of the Sun.  The
catalog is based on a search of the astronomical literature since 1988.  This search
produced information on well over 100 clusters, most of which were identified in various
systematic surveys (e.g, Lada et al.  1991b; Hodapp 1994:  Carpenter Heyer \& Snell 2000).
From this list we selected 76 clusters which met the following criteria:  1) evidence for
embedded nature by association with a molecular cloud, HII region or some significant
degree of optical obscuration or infrared extinction, 2) identification of 35 or more
members above field star background within the cluster field, and 3) location within $\sim$
2 Kpc of the Sun.  Due to large distance uncertainties of regions slightly beyond 2.0 Kpc,
such as the W3 molecular clouds ( 1.8--2.4 Kpc) we have included clusters with published
distance estimates of up to 2.4 Kpc.  Our catalog of nearby embedded clusters is
presented in Table 1 which lists the cluster name, approximate location,
distance, radius, number of members, and absolute magnitude limits of the corresponding
imaging observations.  These data were compiled from the references listed in the last
column.

Given the heterogeneous nature of the observations from which this sample is drawn, this
catalog cannot be considered complete.  In particular, southern hemisphere regions, such as
the Vela complex are not well represented since little observational data exists for this
portion of the Galaxy.  In addition at least 24 additional clusters have been identified in
the Rosette GMC (Phelps \& Lada 1997), North American and Pelican nebula (Cambresy, Beichman
\& Cutri 2002) and Cygnus X region (Dutra \& Bica 2001) but no properties have been presented
in the literature for them and they are not included in our cluster catalog.  Also there is a
general incompleteness for the more distant clusters due to sensitivity limitations.  We
estimate later, that this 2 Kpc sample is complete to only factors of 3-4.  Although not
complete, the catalog is, however, likely representative of the basic statistical properties
of embedded clusters within $\sim$ 2 Kpc.  This is because a significant portion of the
catalog is comprised of clusters drawn from systematic imaging surveys of individual cloud
complexes and a survey (Hodapp 1994) which is reasonably complete for clusters associated
with outflows, a primary tracer of very recent star formation activity in molecular clouds.
Moreover, the subset of clusters found in systematic surveys of nearby GMCs (Orion, Monoceros
and Perseus) is also likely to be reasonably complete for clusters with 35 or more members.


\begin{figure} 
\centerline{\psfig{figure=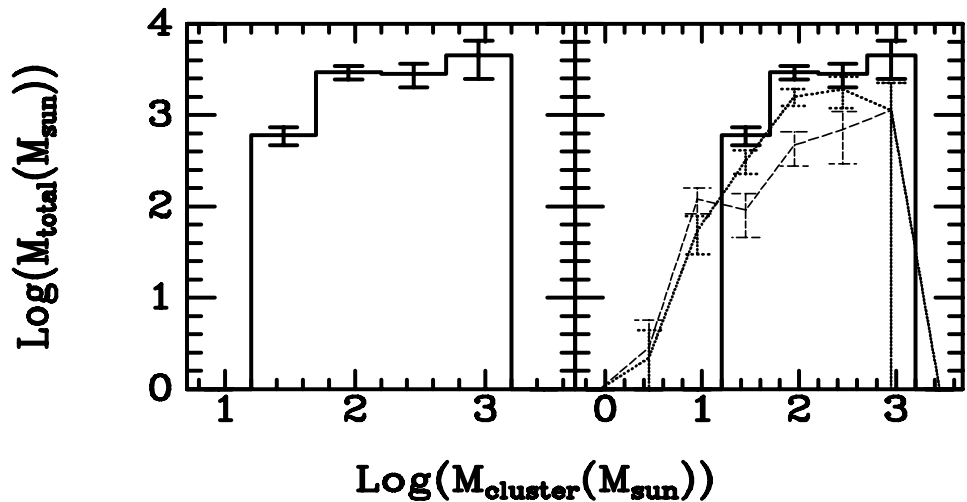,width=5.8in}}
\caption{The embedded cluster mass distribution function (ECMDF)
for the entire embedded cluster catalog is displayed in the left panel. 
This plot traces the distribution of 
total cluster mass (i.e, $N \times M_{EC}$) as a function  of log mass (log($M_{EC}$)). 
The ECMDF is found to be flat for clusters with masses between $\sim$ 50-1000 \msunp. 
This corresponds to an embedded cluster mass function with a spectral index of -2 
(i.e., $dN/dM \propto M^{-2}$). The right panel compares the ECMDF for the entire
catalog with those of two cluster subsamples which are believed to be more complete
at the lowest masses. The dotted line is the Hodapp outflow sample and the dashed
line the sample of all known embedded clusters
within 500 pc of the Sun. The ECMDFs all appear to decline below 50 \msunp.} 
\label{clusmassfn}
\end{figure}

\subsection{The Embedded Cluster Mass Function}

Masses were derived for each cluster in the catalog by assuming a universal IMF (initial
mass function) for all the clusters.  We adopted the IMF of the Trapezium cluster
that was derived by Muench et al.
(2002) from modeling of the cluster's K-band luminosity function (KLF).  We then used the
KLF models of Muench et al.  (2002) to predict infrared source counts as a function of
differing limiting magnitudes for two model clusters whose ages correspond to that of the
Trapezium (0.8 Myr) and IC 348 (2 Myr) clusters.  This was necessary to attempt to account
for the expected luminosity evolution of the PMS populations of embedded clusters (see
discussion below, Section 4).  A conversion factor from total source counts (for a given
limiting magnitude) to total mass was then determined for each synthetic cluster.  The
infrared source counts listed for each observed cluster in the catalog were adjusted for
distance and variable detection limits and then directly compared with the two model
predictions.  In most cases the near-IR limits are faint enough (i.e., the IMF is
reasonably sampled) that the both models yielded cluster masses that agree extremely well
for the two different ages.  Given that no age information is available for the bulk of the
clusters in the catalog, we adopted a conversion factor that was the average of the
Trapezium and the IC 348 cluster ages.  Additionally, we assumed that all clusters have an
average extinction of 0.5 magnitudes in the K band. The masses we have derived 
are probably uncertain to less than a factor of 2 for most clusters.

The derived cluster masses in our sample range from about 20 to 1100 \msunp.  In
Figure~\ref{clusmassfn} (left panel) we present the embedded cluster mass distribution function (ECMDF)
for all the clusters in our sample.  The ECMDF was derived by summing individual embedded
cluster masses ($M_{ec}$) in evenly spaced logarithmic mass bins, 0.5 dex in width,
beginning at Log($M_{ec}$)=1.2.  (The boundaries of the bins were selected to insure
that the least populated bin would have more than one object.)  The ECMDF is equal to
$M_{ec} \times$ dN/dlog$M_{ec}$, and thus differs by a factor of $M_{ec}$ from the mass
function (dN/dlog$M_{ec}$) of embedded clusters.

The histogram with the solid line represents the ECMDF for the entire cluster catalog (i.e.
for clusters having N$_*>$35 and D $<$ 2.4 Kpc).  The mass distribution function displays
two potentially significant features.  First, the function is relatively flat over a range
spanning at least an order of magnitude in cluster mass (i.e., 50 $\leq M_{ec} \leq$ 1000
\msunp).  This indicates that, even though rare, 1000 \msun clusters contribute a
significant fraction of the total stellar mass, the same as for the more numerous 50--100
\msun clusters.  Moreover, more than 90\% of the stars in clusters are found in clusters
with masses in excess of 50 \msun corresponding to populations in excess of 100 members.
The flat mass distribution corresponds to an embedded cluster mass spectrum (dN/d$M_{ec}$)
with a spectral index of -2 over the same range.  This value is quite similar to the
spectral index (-1.7) typically derived for the mass spectrum of dense molecular cloud
cores (e.g.  Lada, Bally \& Stark.  1991c).  The fact that the embedded cluster mass
spectrum closely resembles that of dense cloud cores is very interesting and perhaps
suggests that a uniform star formation efficiency characterizes most cluster forming dense
cores.  The index for the mass spectrum of embedded clusters is also essentially the same
as that (-1.5 to -2) of classical open clusters (e.g., van den Berg \& Lafontaine 1984;
Elmegreen \& Efremov 1997).

The second important feature in the ECMDF is the apparent drop off in the lowest mass bin
($\sim$ 20-50 \msunp).  Given that our cluster catalog only included clusters with more
than 35 stars, it is likely that we will be considerably more incomplete for clusters in
the 20 to 50 M$_\odot$\ range than for the higher mass clusters.  To test the significance
of this fall off to low cluster masses we consider the mass distribution function of a
subset of clusters drawn from a sample of local clouds where observations are reasonably
complete.  These were selected from systematic large scale NIR surveys of 4 molecular
clouds (L1630, L1641, Perseus and Mon R2) without applying any lower limit to the size of
the cluster population.  Therefore, this sample should be sensitive to the full mass range
of clusters in these representative GMCs.  This local molecular cloud sample is plotted as
a dashed line in the right panel of Figure~\ref{clusmassfn}.  While the statistical errors
due to the small sample size are large, the local sample confirms that there is indeed a
drop off in total cluster mass for the lowest mass clusters.  As a further check, we also
plot the ECMDF for the Hodapp (1994) sample, again without applying any lower limit to the
richness of the cluster.  We choose the Hodapp sample since it is selected from a complete
sample of outflows, indicative of very young stellar objects, and should not be biased to
any particular mass range of clusters.  All three samples are consistent with a fall off in
the cluster mass spectrum below about 50 M$_\odot$.  Even if the cluster samples are not
formally complete they should be representative of the total local cluster population
within 2 Kpc.  Therefore, we conclude that the drop off in the ECMDF at masses less than 50
\msun is significant.  Consequently, there appears to be a characteristic cluster mass (50
M$_\odot$) above which the bulk of the star forming activity in clusters is occurring.
Recently, Adams and Myers (2000) suggested, based on dynamical modeling of open clusters
and knowledge of the cluster formation rate, that most clustered star formation occurs in
clusters with between 10 and 100 stars.  However, our results imply that no more than about
10\% of all stars are formed in such small clusters.  The discrepancy results from Adams \&
Myers use of the Battinelli \& Capuzzo-Dolcetta (1991) catalog of open clusters which
undercounts clusters with ages less than 3 Myrs and underestimates the cluster formation
rate as discussed below.

Using the masses in Table 1 we can estimate the contribution to the star
formation rate made by embedded clusters.  Because of the incompleteness of our sample,
this estimate necessarily will be a lower limit.  To minimize the effect of incompleteness
we can calculate the star formation rate for the local (d $<$ 500 pc) subset of clusters
for which we are likely to be reasonably complete.  For this subsample we calculate a
local star formation rate of $\geq$ 1-3$\times$10$^{-9}$ M$_\odot$ yr$^{-1}$ pc$^{-2}$ assuming
typical embedded cluster ages of $\sim$1-2 Myrs.  This rate is in reasonable agreement with the
local star formation rate derived from field stars by Miller and Scalo (1979) of between
3-7 $\times$10$^{-9}$ M$_\odot$ yr$^{-1}$ pc$^{-2}$.  This suggests that embedded clusters
may account for a large fraction of all star formation occurring locally as has been
suggested by other considerations (Lada et al.  1991b; Carpenter 2000).  Extending our
sample to 1 and 2 Kpc gives star formation rates of 1-0.7 $\times$10$^{-9}$ M$_\odot$
yr$^{-1}$ pc$^{-2}$, respectively (for $\tau_{age} \sim$ 1 Myr).  The systematic drop in the star formation rate with distance likely
reflects progressively more incomplete cluster surveys as we move to greater distances.
If we assume that we are nearly complete for the local 0.5 Kpc sample, then the drop in
the calculated birthrates would imply that we are incomplete by factors of at least 3 to 4
for the 1 to 2 Kpc samples.

\subsection{Birthrates and Star Formation}

The embedded cluster catalog can be used to estimate a lower limit to the birthrate of
embedded clusters in molecular clouds.  Early estimates of the embedded cluster birthrate,
based primarily on the number of clusters in the Orion cloud complex, found the rate to be
extremely high compared to the birthrate of classical open clusters suggesting that only a
small fraction of embedded clusters survived emergence from molecular clouds to become
classical open clusters (Lada \& Lada 1991).  Our more extensive embedded cluster catalog
with an order of magnitude more clusters, allows for a straightforward but much more
meaningful estimate of this important formation rate.  For (53) clusters within 2.0 Kpc we
estimate the formation rate to be between 2-4 clusters Myr$^{-1}$ Kpc$^{-2}$ for assumed
average embedded cluster ages of 2 and 1 Myrs, respectively.  Although this rate is
a lower limit, it is a factor of
8--16 times that (0.25 Myr$^{-1}$ Kpc$^{-2}$) estimated for classical open clusters by
Elmegreen \& Clemens (1985) and 5-9 times that (0.45 Myr$^{-1}$ Kpc$^{-2}$) estimated by
Battinelli \& Capuzzo-Dolcetta (1991) for a more complete open cluster sample within 2 Kpc
of the Sun.  This difference in birthrates between embedded and open clusters
represents an enormous discrepancy and is of
fundamental significance for understanding cluster formation and evolution.

\begin{figure} 
\centerline{\psfig{figure=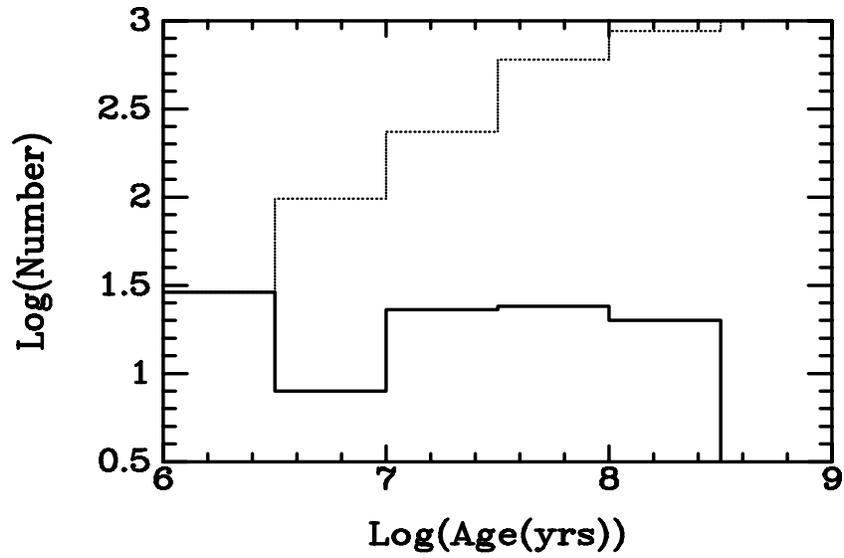,width=5in}}
\caption{Observed frequency distribution of ages for  open and embedded clusters
within 2 Kpc of the Sun (solid line) compared to that (dotted line) predicted for a
constant rate of star formation adjusted for cluster luminosity evolution. 
All embedded clusters fall into the first bin. 
The large discrepancy between the predicted and observed numbers indicates a high infant 
mortality rate for protoclusters. } 
\label{clustages} 
\end{figure}

By combining our embedded cluster catalog with the open cluster catalog of Battinelli \&
Capuzzo-Dolcetta (1991) we can examine the age distribution of all clusters, open and
embedded, within 2 Kpc of the Sun.  The Battinelli \& Cappuzzo-Dolcetta catalog contains
about 100 classical open clusters and is thought to be complete out to a distance of 2 Kpc
from the Sun for clusters with M$_V <$~-4.5.  In Figure~\ref{clustages} we plot the
distribution of ages of all known clusters both embedded and open within 2 Kpc.  Embedded
clusters populate the lowest age bin.  We have included only those embedded clusters with
masses greater than 150 \msun to correspond to the magnitude-limited selection of
Battinelli \& Cappuzzo-Dolcetta.  This represents roughly one-third of our sample of
clusters with published distances of 2~Kpc or less.  The average mass of these embedded
clusters is 500 \msunp, the same as that estimated for the open cluster sample by
Battinelli \& Cappuzzo-Dolcetta (1991).  The number of clusters is found to be roughly
constant as a function of age for at least 100 Myr.  In Figure~\ref{clustages} we also
compare the merged cluster age distribution with the expected age distribution for a
constant rate of cluster formation.  Our prediction also includes an adjustment for the
expected luminosity fading of clusters below the detection limits following the
perscription of Battinelli \& Cappuzzo-Dolcetta (1991).  There is a large and increasing
discrepancy between the expected and observed numbers.  These distributions clearly confirm
earlier speculations that the vast majority of embedded clusters do not survive emergence
from molecular clouds as identifiable systems for periods even as long as 10 Myr.  {\it
Figure~\ref{clustages} suggests an extremely high infant mortality rate for clusters.}
Less than $\sim$ 4\% of the clusters formed in molecular clouds are able to reach ages
beyond 100 Myr in the solar neighborhood, less than 10\% survive longer than 10 Myr.
Indeed, most clusters may dissolve well before they reach an age of 10 Myr.  If we consider
our entire sample of embedded clusters we predict, after similarly adjusing for fading,
that at least 4100 clusters would be detected within 2 Kpc of the Sun with ages less than
300 Myr.  The WEBDA open cluster catalog lists roughly 300 open clusters of this age or
less within 2 Kpc, suggesting that only about 7\% of all embedded clusters survive to
Pleiades age.  It is likely that only the most massive clusters in our catalog are
candidates for long term survival.  Roughly 7\% of embedded clusters in our catalog have
masses in excess of 500 \msunp, and this likely represents a lower limit to the mass of an
embedded cluster that can evolve to a Pleiades-like system.  Moreover,
Figure~\ref{clustages} also indicates that the disruption rate for bound clusters between
10--100 Myrs of age is significant, probably due to encounters with GMCs.  Many of the
observed open clusters in this age range may also not be presently bound (Battinelli \&
Capuzzo-Dolcetta 1991).

The discovery of large numbers of embedded clusters coupled with the high birthrates and
star formation rates we have inferred for them from our analysis of the data in
Table 1, suggests that such clusters may account for a significant fraction of
all star formation in the Galaxy.  However, because of the incompleteness of our sample it
is difficult to produce an accurate estimate of the actual fraction of stars born in
embedded clusters from statistical analysis of the data in our catalog.  The best
estimates of this quantity are derived from systematic, large scale surveys of individual
GMCs.  The first systematic attempt to obtain an inventory of high and low mass YSOs in a
single GMC was made by Lada et al.  (1991b) who performed an extensive near-infrared
imaging survey of the central regions (~ 1 square degree) of the L1630 GMC in Orion.
Their survey produced the unexpected result that the vast majority (60-90\%) of the YSOs
and star formation in that cloud occurred within a few (3) rich clusters with little
activity in the vast molecular cloud regions outside these clusters.  A subsequent survey by
Carpenter (2000) using the 2MASS database to investigate the distribution of young stars
in 4 nearby molecular clouds, including L1630 produced similar results with estimates of
50-100\% of the clouds' embedded populations be confined to embedded clusters.  In both
studies the lower limits were derived with no correction for field star contamination
which is substantial.  Consequently, it is likely that the fraction of stars formed in
clusters is very high (70-90\%).  Subsequent near-infrared surveys of L1630 (Li, Evans \&
Lada 1997) as well as other molecular clouds such as Mon OB1 (Lada, Young \& Greene 1993),
the Rosette (Phelps \& Lada 1997) and Gem OB1 (Carpenter, Snell \& Schloerb 1995) have
yielded similar findings suggesting that formation in clusters may be the dominant mode of
formation for stars of all masses in GMCs and that embedded clusters may be the
fundamental units of star formation in GMCs.  Since GMCs account for almost all star
formation in the Galaxy, most field stars in the Galactic disk may also have originated in
embedded clusters.

\subsection{Association with Molecular Gas and Dust}

The intimate physical association with interstellar gas and dust is the defining
characteristic of embedded clusters.  Embedded clusters can either be partially (i.e.,
A$_V$ $\sim$ 1--5 mag.)  or deeply (i.e., A$_V$ $\sim$ 5--100 mag) immersed in cold dense
molecular material or hot dusty HII regions.  The degree of their embeddedness in molecular
gas is related to their evolutionary state.  The least evolved and youngest embedded
clusters (e.g., NGC 2024, NGC 1333, Ophiuchi, MonR2, and Serpens) are found in massive
dense molecular cores, while the most evolved (e.g., the Trapezium, NGC 3603, IC 348)
within HII regions and reflection nebulae or at the edge of molecular clouds.  Our present
understanding of the relation of dense cores and embedded clusters is largely guided by the
coordinated surveys of such clouds as L1630 (Orion B), Gem OB1 and the Rosette (Mon OB2).
These are the clouds for which the most systematic and complete surveys for both embedded
clusters and dense molecular material exist (Lada 1992; Carpenter, Snell \& Schloerb 1996;
Phelps and Lada 1997).  These studies all show that embedded clusters are physically
associated with the most massive (100--1000 \msunp) and dense (n(H$_2$) $\sim$ 10$^{4-5}$
cm$^{-3}$) cores within the clouds.  These cores have sizes (diameters) typically on the
order of 0.5-1 pc.  The typical star formation efficiencies range between 10-30\% for these
systems.  The gas densities correspond to mass densities of 10$^{3-4}$ \msun pc$^{-3}$
suggesting that clusters with central densities of a few times 10$^3$ \msun pc$^{-3}$ can
readily form from them.

Typically less than 10\% of the area and mass of a GMC is in the form of dense gas.  This
gas is non-uniformly distributed through the cloud within numerous discrete and localized
cores.  These cores range in size between about 0.1 - 2 pc and in mass between a few solar
masses to up to a thousand solar masses.  The largest cores which spawn clusters are highly
localized and occupy only a very small fraction (a few \%) of the area of a GMC.  Numerous
studies have indicated that the mass spectrum ($dN \over dm$) of dense molecular cloud
cores is a power-law with an index of $\alpha$ $\sim$ -1.7 (e.g, Lada, Bally \& Stark 1991,
Blitz 1993, Kramer et al.  1998).  For such a power-law index, most of the mass of dense
gas in a cloud will be found in its most massive cores, even though low mass cores
outnumber high mass cores.  Stars form in dense gas and it is not surprising therefore that
a high fraction of all stars form in highly localized rich clusters, since most of a
cloud's dense gas is contained in its localized massive cores.  Moreover, as discussed
earlier, the mass spectrum of cores is very similar to that of both embedded and classical
open clusters.

Not all massive dense cores in molecular clouds are presently forming clusters (e.g., Lada
1992).  However, in the L1630 cloud, the cores with clusters appear to contain more gas at
very high density (n(H$_2$) $> 10^5$ cm$^{-3}$) and to be more highly clumped or structured
than those cores without clusters (Lada, Evans \& Falgarone 1997).  Whether this difference
in physical properties is a cause or a result of the formation of a cluster in a massive
core is unclear.  Studies of the distribution of dust continuum emission in the Ophiuchi
(Motte, Andre \& Neri 1998), Serpens (Testi \& Sargent 1998) and the NGC 2068/2071 (Motte
et al.  2001) cluster forming cores reveal numerous small scale ($\sim$ 5000 AU) clumps
whose mass spectra are characterized by power-law slopes steeper than those of cloud cores
but very similar to those which characterize the stellar IMF (see below).  This would
suggest that there is a direct mapping of clump mass to stellar mass and that the substructure
of cluster forming cores reflects the initial conditions of the star formation process in
dense cores.  On the other hand, detailed observations of the NGC 1333 cluster forming core
paint a very different picture.  Here the density structure appears to be defined by
numerous shells and cavities associated with intense outflow activity (Sandell \& Knee
2001; Lefloch et al.  1998).  This suggests that much of the structure in the core is the
result of excavation by outflows from the young stars themselves and is a post- rather than
pre- star formation condition in this and perhaps other cores where outflow activity is
occurring.  In this regard it would certainly be interesting to obtain high resolution dust
continuum maps of the massive cores that are not forming clusters to see if the mass
spectra of pre-stellar clumps in cores for which active star formation has not yet taken
place is similar to that in cores in which star formation is already under way.  A
significant difference could implicate outflow activity as an agent that transforms an
initial cloud-like clump mass spectrum into a clump spectrum more similar to that of stars.

\subsection{Internal Structure and Mass Segregation}

The structure of an embedded cluster is of great interest since it likely possesses the
imprint of the physical process responsible for its creation.  In particular, structure in
the youngest embedded clusters reflects the underlying structure in the dense molecular gas
from which they  formed.  Although all embedded clusters appear to display structure at some
level, they can be characterized by two basic structural types: 1)  hierarchical-type clusters
exhibit surface density distributions with multiple peaks and often significant structure over a
large range of spatial scale. 2) centrally condensed-type embedded clusters exhibit highly
concentrated surface density distributions with relatively smooth radial profiles which can
be described to a good approximation by simple power-law functions (e.g., $\rho_*(r) \sim
r^{-q}$) or King-like (isothermal) potentials.  In this sense they are similar to classical
open clusters. The relative frequency of these two types of structure in clusters
is presently unknown, although there are clear examples of each in the literature.

Examples of hierarchical-type clusters include the deeply embedded double cluster NGC 1333
(Lada, Alves \& Lada 1996) and the partially embedded cluster NGC 2264 which is highly
structured (Lada, Green \& Young 1992, Piche 1993).  Figure~\ref{n2264map} shows a map of
the spatial distribution of infrared sources in the NGC 2264 cluster constructed from the
data of Lada, Greene \& Young (1992).  The cluster appears to be a double-double or
quadruple cluster containing at least two levels of hierarchy in its spatial structure.
The existence of hierarchical structure over large scales in star forming regions has been
well documented and is thought to be a signature of the turbulent nature of the
interstellar gas and dust out of which GMCs, their dense cores and
ultimately stars form (e.g., Elmegreen et al.  2000). 

\begin{figure} 
\vskip 6.0in

\epsfxsize=5.8in 
\caption{Contour map of the surface density of J-band infrared sources in
the partially embedded cluster NGC 2264. This is an example of a 
cluster that displays a hierarchical structure.} 
\label{n2264map} 
\end{figure}

The Trapezium-ONC cluster (Hillenbrand \& Hartmann 1998), IC 348 (Lada \& Lada 1995; Muench
et al 2002), NGC 2024, NGC 2071 (Lada et al.  1991b) and NGC 2282 (Horner, Lada \& Lada
1997) have been shown to have strong central concentrations and radial surface density
profiles that can typically be fit by simple power-laws ($q \approx 1.5$) as well as King
models.  Such structure is a signature of the global dominance of gravity (over turbulence,
for example) in the formation of these systems.  Whether this structure is a primordial
property of these clusters or a result of evolution from an initially more structured and
hierarchical state is not clear.  It is interesting to note in this context that each of
these clusters is about the spatial extent of a single subcluster in NGC 2264.  Overall,
the centrally condensed clusters do exhibit some structure, but it is considerably less
dramatic and more subtle than that observed in clusters such as NGC 2264.  For example,
Lada \& Lada (1995) documented a handful of small, satellite subclusters in the outer
regions of IC 348.  However, given the small numbers of stars within them, one cannot rule
out the hypothesis that these structures are merely the expected spatial fluctuations in
the overall power-law fall off of the cluster's radial density distribution.  Images of the
Trapezium cluster in L band (3.4 $\mu$m) revealed that about 10\% of the sources in the
cluster belong to a deeply embedded population, whose surface density distribution differs
from that of the main (less buried) cluster (Lada et al.  2000).  In particular, the
surface density distribution of the embedded population has a different orientation being
more closely aligned with the ridge of molecular gas and dust at the back of the cluster.
This population is also not nearly as centrally condensed.  Another indication of subtle
structure in the Trapezium-ONC and IC 348 clusters is the evidence for a spatial variation
in the clusters' mass functions, both of which appear to exhibit an excess of the lowest
mass stars in their outer regions perhaps suggestive of some degree of mass segregation
(Hillenbrand \& Carpenter 2001; Muench et al.  2002).

The question of mass segregation in embedded clusters is of great interest.  Evidence for
mass segregation in open clusters is well documented (e.g., Elmegreen et al.  2000) and
likely due to dynamical evolution and equipartition of energy in those systems.  But is
mass segregation also a primordial property of clusters or only achieved after significant
dynamical evolution?  Bonnell \& Melvyn (1998) have argued that embedded clusters, like the
Trapezium, are too young to have dynamically evolved significant mass segregation.  If
there is mass segregation in this  cluster it must be imprinted by the star formation
process.  The strongest indication of possible mass segregation in embedded clusters
derives from observations which suggest that the most massive stars in some of these
systems are preferentially found near the cluster centers.  This phenomenon has been
observed in the Trapezium (Hillenbrand \& Hartmann 1998), NGC 2071 and NGC 2024 (Lada et
al.  1991b).  Nurnberger \& Petr-Gotzens (2002) found a steepening of the mass function in
the outer region of NGC 3603 and Jiang et al.  (2002) found an exponential, rather than
power-law, radial decline of massive OB stars in M17 suggesting perhaps more extensive mass
segregation in these more massive clusters.  In other embedded clusters, such as MonR2, no
evidence has been found for any significant mass segregation (Carpenter et al.  1997).
Indeed, there are even examples where the high mass stars are found both in the outer
regions as well as the central regions of a cluster (e.g., IC 5146; Herbig \& Dahm 2002).
In this context it is interesting to note that a study of massive young clusters in 
the LMC also found little evidence for mass segregation (Elson, Fall \& Freeman 1987).
The extent of the phenomenon of mass segregation in embedded clusters remains far from
clear.  Unfortunately, the inherent uncertainties due to small number statistics make it
difficult to definitively investigate this issue in all but the most rich and massive
clusters, which themselves are rare and typically very distant.

\subsection{Ages and Age Spreads}

The ages and age spreads of embedded clusters and their members are fundamental parameters
which are among the most uncertain and difficult to determine for such young systems.
Knowledge of these two timescales are critical for understanding the evolutionary
appearance and state of a cluster and its star formation history.  For example, the ratio
of cluster age, $\tau_{age}$, to such timescales as the crossing time ($\tau_{cross}$), the
relaxation time, ($\tau_{r}$), the evaporation time ($\tau_{ev}$), etc.  determines the
dynamical state of the cluster.  The relation between $\tau_{age}$ and the various
timescales of early stellar evolution determine the evolutionary demographics of cluster
members (e.g., the number of cluster members that are protostars, or PMS stars with and
without disks, etc.).  Indeed, as will be discussed later, the protostellar, $\tau_{ps}$,
and disk, $\tau_{disk}$, evolution timescales can be inferred from knowledge of
$\tau_{age}$ and an observational census of either the protostars or disk bearing stars
(respectively) within a cluster, or a sample of clusters of varying age.  The age spread,
$\Delta\tau_{sf}$, gives the duration of star formation or the gestation timescale for the
cluster population.  Also of interest is the star formation rate which is essentially given
by the ratio of the number of cluster members to the gestation timescale (i.e.,
$N_*/\Delta\tau_{sf})$ .  These latter timescales provide important constraints for
understanding the physical process of star formation within the cluster.  Finally, embedded
cluster ages are critical for dating molecular clouds, since cloud ages cannot be
determined from observations of the dust and gas within them.  The age of an embedded
cluster provides an interesting lower limit to the age of the molecular cloud from which it
formed.  Indeed, the relative absence of molecular emission from around clusters with ages
in excess of 5 Myr, has long suggested that the lifetimes of molecular clouds typically do
not exceed 5-10 Myr (Leisawitz et al 1989).

The most reliable method for age dating clusters and their members is through use of the
Hertzberg-Russell diagram (HRD) where the positions of member stars are compared with the
locations of theoretical PMS evolutionary tracks.  Unfortunately, the theoretical
trajectories of PMS stars on the HRD can be highly uncertain, particularly for cluster ages
of 1 Myr or less and for low mass stars and substellar objects (e.g, Baraffe et al.  2002).
In addition, empirical measurements of the two stellar parameters (i.e., luminosities and
effective temperatures) that are necessary for placement of individual stars on the HRD
can be very difficult due to such factors as extinction, stellar variability,
binarity, veiling and infrared excess which are common characteristics of PMS stars.
Moreover, often such measurements can only be carried out at infrared wavelengths.  As a
result of these various factors, the ages of embedded clusters are inherently uncertain and
often very poorly constrained by observations.

\begin{figure}
\vskip 6.0in


\caption{The V, V-I color magnitude diagram for IC 348 compiled by 
Herbig (1998). Circles are stars for which spectral types are known,
solid dots indicate stars with H$\alpha$ emission and therefore likely
members. Theoretical isochrones derived from a single set of PMS tracks
are shown and labeled by age (e.g., 3(6) corresponds to 3 Myr.). The
diagram shows that there is an appreciable spread in the locus of 
PMS stars in this cluster corresponding to an apparent age spread which is comparable
to the mean age of the cluster. }  
\label{cmdic348} 
\end{figure}

In practice it is often more straightforward to place stars on the color-magnitude diagram
(CMD) and then transform the theoretical tracks to that observational plane for comparison
and cluster age determinations.  There exist only a small number of embedded clusters, such
as NGC 2024 (Meyer 1996), IC 348 (Herbig, 1998), the Trapezium (Hillenbrand 1997), NGC 2264
(Park et al.  2000), IC 5146 (Herbig \& Dahm 2002) etc., for which sufficient observations
are available to place a significant sample of sources on the CMD and estimate their ages.
Published estimates for the mean ages of these clusters vary between 0.5-3 Myr.
Unfortunately, the uncertainties introduced by using PMS tracks can be on the order of the
derived age of the cluster.  For example, Park et al.  (2000) derived the age of NGC 2264
using 4 different PMS models with resulting values of 0.9, 2.1, 2.7, and 4.3 Myrs for the
age of the cluster.  Another problem is that an individual PMS model can give different
ages for high and low mass stars in the same cluster (e.g., Hillenbrand 1997).  On the
other hand, the relative mean ages of young clusters can be established to much greater
precision by using a single or consistent set of PMS models to extract the cluster ages
(e.g., Haisch et al.  2001a).

Published comparisons of embedded cluster HRDs or CMDs with theoretical PMS tracks indicate
age spreads that are usually of the same order or even greater than the mean cluster ages.
For example, age spreads of 5.5, 8.0, 10.0 and 15.3 Myr were found by Park et al.  for NGC
2264 using four different PMS models for the 2-3 Myr old cluster.  A good example of an
embedded cluster CMD is shown in Figure~\ref{cmdic348} which displays the CMD for the IC
348 cluster constructed by Herbig (1998).  The positions of suspected cluster members form
a well defined, but relatively wide, PMS.  Comparison with the isochrones derived from a
single set of PMS tracks gives a cluster mean age of $\sim$ 2 Myr and an age spread of
approximately 5 Myrs.  In principle, the detailed distribution of stars within the PMS band
in the plane of the CMD reflects the star formation history of the embedded population.
Indeed, from a detailed examination of the distribution of stars in the HRDs of a
number of embedded clusters (including the Trapezium, IC 348, NGC 2264 and Rho Ophiuchi),
Palla and Stahler (2000) produced intriguing evidence for a strongly time dependent star
formation rate in these regions.  Using a consistent analysis and a single set of PMS
tracks, they found that star formation appeared to be accelerating with time, with the star
formation rate reaching its peak in the last 1-2 Myr in all clusters still associated with
significant molecular gas.  

However, it is difficult to evaluate the significance of age
spreads and distributions estimated from CMDs of embedded clusters because differential
extinction, source variability, infrared excess, binarity, and contamination by field stars
can contribute significantly to the intrinsic scatter in the diagram (e.g., Hartmann 2001).
The uncertainties due to such factors as variability, infrared excess and extinction are
expected to be greater for younger clusters.  Figure \ref{cmdn2362} shows the CMD obtained
for NGC 2362, a 5 Myr old, exposed, open cluster where such uncertainties should be
minimized (Moitinho et al.  2001).  The PMS of this cluster is very well defined and
relatively narrow indicating a clear upper limit to its age spread of $<$ 3 Myr.  In this
cluster, where the total and differential extinction are barely measurable, and stellar
activity associated with the youngest stars minimal, the CMD indicates a simple star
formation history characterized by cluster formation in a rapid, coeval burst of activity
less than 3 Myr in duration.  These observations also may suggest that a significant
portion of the observed scatter in the CMDs of other younger and embedded clusters is due
to factors other than age.  Unfortunately, it is presently not possible to determine
whether the large spreads in embedded cluster CMDs result from a wide variety of gestation
times, accelerating star formation or other factors.  Since the number of embedded clusters
with age determinations is small, a systematic and detailed examination of a larger sample
of embedded and young open cluster CMDs would be useful in resolving this issue.  At
present, self-consistent determinations of the mean ages and, in particular, the relative
mean ages of clusters may be the most robust information about star formation histories
that can be extracted from CMD or HRD analysis.

\begin{figure}
\vskip 6.0in

\caption{The V, V-I color magnitude diagram for the 5 Myr open
cluster NGC 2362 (Moitinho et al 2001). The left panel shows the CMD for
all stars in the NGC 2362 field. A narrow, well defined PMS band is evident stretching 
over  a range of at least 9 magnitudes in V, from A stars to M stars near the hydrogen
burning limit. The right panel shows a contour
plot of difference between the surface density of stars in the CMD of the cluster (on) and
that of stars in the CMD of a nearby control field (off). The locus of points corresponding to
the ZAMS and a 5 Myr PMS isochrone are also plotted for comparison. The narrow width of
the cluster's PMS indicates that the cluster formed in a coeval burst of star formation
less than 3 Myr in duration. Figure courtesy of J. Alves. }  
\label{cmdn2362} 
\end{figure}

\section{EMBEDDED CLUSTERS AND THE INITIAL MASS FUNCTION}

\subsection{Background}

A fundamental consequence of the theory of stellar structure and evolution is that, once
formed, the subsequent life history of a star is essentially predetermined by one
parameter, its birth mass.  Consequently, detailed knowledge of the initial distribution
of stellar masses at birth (i.e., the IMF) and how this quantity varies through time and
space is necessary to predict and understand the evolution of stellar systems, such as
galaxies and clusters.  Detailed knowledge of the IMF and its spatial and temporal
variations is also particularly important for understanding the process of star
formation, since it is the mysterious physics of this process that controls the
conversion of interstellar matter into stars.  Unfortunately, stellar evolution theory
is unable to predict the form of the IMF.  This quantity must be derived from
observations.  However this is not a straightforward exercise, since stellar mass is not
itself an observable quantity.  Stellar radiant flux or luminosity is the most readily
observed property of a star.  Determination of stellar masses therefore requires a
transformation of stellar luminosities into stellar masses which in turn requires
knowledge of stellar evolutionary states.

Numerous techniques have been employed in an attempt to determine the IMF both for the
galactic field star population and in open clusters.  These techniques and results have
been extensively reviewed in the literature (e.g., Scalo 1978, 1986; Gilmore \& Howell
1998; Meyer et al.  2000; Kroupa 2002).  IMFs derived from these studies appear to
exhibit two similar general properties.  First, for stars more massive than the sun the
IMF has a nearly power-law form with the number of stars increasing as the stellar mass
decreases.  If we adopt the classical definition that the IMF ($\xi(log m_*)$) is the
number of stars formed per unit volume per unit {\it logarithmic} mass interval, the
slope at any point is then:  $\beta \equiv \partial log \xi (logm_*) / \partial logm_*$,
and $\beta \approx$ --1.3 for masses greater than one solar mass (e.g., Massey 1998).
This is very similar to the value (--1.35) originally derived for field stars by
Salpeter (1955).  Second, the IMF breaks and flattens near but slightly below 1 \msun,
departing significantly from a Salpeter slope.  At the lowest masses (i.e., 0.5 - 0.1
\msun), however, there is considerable debate concerning whether the IMF declines, rises
or is flat and whether or not it extends smoothly below the hydrogen burning limit (HBL)
to substellar masses.

However, IMF determinations for local field stars and in open clusters are hampered by a
number of serious difficulties.  To deduce the IMF for field stars requires compilation of
a volume limited sample of nearby stars.  This in turn requires accurate distance
measurements, usually parallaxes, for all stars in the sample.  To obtain the necessary
complete sample to as low a mass as possible, necessitates that this volume be limited to
stars relatively nearby the sun (d $\sim$ 5-25 pc), because of the extreme faintness of the
lowest mass stars and the limitations inherent in the distance determinations.  Such
samples suffer from incompletness for both the highest mass stars, due to their rarity and
complete absence in the solar neighborhood, and the lowest mass stars due to their
faintness.  Moreover, such samples contain stars formed over a time interval encompassing
billions of years (essentially the age of the galactic disk).  Therefore, the mass function
derived directly from observations of field stars is a present day mass function (PDMF) and
must be corrected for the loss of higher mass stars due to stellar evolution in order to
derive the IMF of the sample.  This, in turn, requires the assumptions of both a star
formation rate, usually taken to be constant, and a time independent functional form of the
IMF.  The standard final product is an IMF which is time-averaged over the age of the Milky
Way disk.  Details regarding any dependence of the IMF on either space or time over the
history of the Galaxy are necessarily lost in the time-averaged IMFs derived from nearby
field stars.  Finally, it is very difficult to measure the IMF below the HBL from
magnitude-limited studies of the field.  This is because the luminosities of brown dwarfs
continue to fade throughout their entire life history, so the number of substellar objects
at any brightness is always a time-dependent mixture of brown dwarfs of varying mass and
age and it depends sensitively on the formation history of brown dwarfs in the Galactic
disk.

Stellar clusters have played an important role in IMF studies because they present
equidistant and coeval populations of stars of similar chemical composition.  Compared
to the disk population, clusters provide an instantaneous sampling of the IMF at
different epochs in galactic history (corresponding to the different cluster ages) and
in different, relatively small volumes of space.  This enables investigation of possible
spatial and temporal variations in the IMF.  However some of these advantages are
mitigated by the larger distances of visible open clusters (compared to local field
stars) which reduces the sensitivity to faint low mass stars and by field star
contamination which seriously hampers determination of cluster membership and
achievement of completeness, especially at low masses.  In addition dynamical evolution
produces both mass segregation and evaporation and depletes the low mass population of
clusters requiring uncertain corrections to be applied in order to obtain their IMFs.
Also, stellar evolution depletes the high mass end of the IMF in clusters older than ten
million years or so and thus must be accounted for as well.

Using young embedded clusters for IMF determinations alleviates many of these issues.
For example, embedded clusters are often significantly more compact than visible open
clusters minimizing field star contamination, except at the very lowest masses.
Moreover, the molecular gas and dust associated with such clusters can screen background
stars even at faint magnitudes further reducing background contamination and associated
difficulties with membership determinations.  In addition, {\it embedded clusters are
too young to have lost significant numbers of stars due to stellar evolution or
dynamical evaporation, thus their present day mass functions are, to a very good
approximation, their initial mass functions}.  Embedded clusters are also particularly
well suited for determining the nature of the IMF for low mass stars and substellar
objects.  This is because low mass stars in embedded clusters are primarily pre-main
sequence stars, and thus are brighter than at any other time in their lives prior to
their evolution off the main-sequence.  At these young ages, substellar objects or brown
dwarfs are also significantly more luminous than at any other time in their subsequent
evolution, and moreover have brightnesses comparable to the lowest mass stars.  Indeed,
infrared observations of modest depth are capable of detecting objects spanning the
entire range of stellar mass from 0.01 to 100 \msun in clusters within 0.5 -- 1.0 Kpc
of the sun. 

However, the study of embedded clusters suffers from two disadvantages:  1) the
clusters are often heavily obscured and cannot be easily observed at optical
wavelengths; 2) the stars in such clusters are mostly pre-main sequence stars and the
timescale for forming them is an appreciable fraction of the cluster age.  Consequently,
uncertain corrections for pre-main sequence evolution and non-coevality must be applied
to the members to derive mass spectra from luminosity functions.  Advances in infrared
detectors have enabled the direct observation of such embedded clusters and helped to
minimize the first disadvantage.  The second disadvantage requires modeling and is more
difficult to overcome (e.g., Zinnecker et al.  1993; Comeron et al.  1993; Fletcher and
Stahler 1994a; Lada \& Lada 1995; Comeron, Rieke \& Rieke 1996; Lada, Lada \& Muench
1998; Luhman et al.  2000; Meyer et al.  2000; Muench, Lada \& Lada 2000; Muench et al.
2002 and others).  Finally, although the underlying mass functions of most embedded
clusters are likely to be a fair representation their initial mass functions, some
embedded clusters can be expected to be in extremely early stages of evolution in which
active star formation is still contributing to building the ultimate cluster IMF.
Appropriate caution must be taken when interpreting the mass functions derived in such
circumstances.

\subsection{Methodology:  From Luminosity to Mass Functions}

Two basic methods have been generally employed to derive mass functions for embedded
clusters.  The first method involves modeling the observed luminosity function of an
embedded cluster to derive the form of its underlying mass function (e.g., Zinnecker et
al.  1993; Fletcher \& Stahler 1994a,b; Lada \& Lada 1995; Megeath 1996; Muench et al.
2000, 2002).  The second method involves use of spectroscopy and/or multi-color
photometry to place individual stars on the HRD.  Comparison of the locations of
these stars in the HRD with the predictions of PMS evolutionary tracks results in
the determination of their individual masses from which the cluster IMF is then directly
constructed (e.g., Hillenbrand 1997, Hillenbrand \& Carpenter 2000).  As will be
discussed below, these methods have their own advantages and disadvantages, but in
general are complementary.  Indeed, in a few studies, IMFs for embedded clusters have
been derived using a combination of these and similar techniques (e.g., Comeron et al
1993, 1996; Luhman et al. 1998, 2000).

\subsubsection{Modelling the Luminosity Function}

Deriving the IMF of an embedded cluster by modelling its luminosity function first
requires the construction of the cluster's luminosity function.  Although determining
the bolometric luminosity function of a cluster would be most desirable for comparison
with theoretical predictions (e.g., Lada \& Wilking 1984; Fletcher \& Stahler 1994a,b),
obtaining the multi-wavelength observations necessary to do so would require prohibitive
amounts of observing time on telescopes both on the ground and in space.  On the other
hand, the monochromatic brightness of a star is its most basic observable property and
infrared cameras enable the simultaneous measurement of the monochromatic brightnesses
of hundreds of stars.  Thus, complete luminosity functions, which span the entire range
of stellar mass, can be readily constructed for embedded stellar clusters with small
investments of telescope time.  The monochromatic (e.g., K band) luminosity function of
a cluster, $dN \over dm_K$, is defined as the number of cluster stars per unit magnitude
interval and is the product of the underlying mass function and the derivative of the
appropriate mass-luminosity relation (MLR):

\begin{equation} 
{dN\over dm_K} = {dN\over dlogM_*} \times {dlogM_*\over dm_K} 
\label{eq1} 
\end{equation}


\noindent
where $m_k$ is the apparent stellar (K) magnitude, and $M_*$ is the stellar mass.  The
first term on the right hand side of the equation is the underlying stellar mass
function and the second term the derivative of the MLR.  With knowledge of the MLR (and
bolometric corrections) this equation can be inverted to derive the underlying mass
function from the observed luminosity function of a cluster whose distance is known.
This method is essentially that originally employed by Salpeter (1955) to derive the
field star IMF.  However, unlike main sequence field stars, PMS stars, which account for
most of the stars in the an embedded cluster, cannot be characterized by a unique MLR.
Indeed, the MLR for PMS stars is a function of time.  Moreover, for embedded clusters
the duration of star formation can be a significant fraction of the cluster's age.
Consequently, to invert the equation and derive the mass function one must model the
luminosity function of the cluster and this requires knowledge of both the star
formation history (i.e., age and age spread) of the cluster as well as the time-varying
PMS mass-luminosity relation.  This presents the two major disadvantages for this
technique.  First, a priori knowledge of the age or star formation history of the
cluster is required and this typically can be derived by placing cluster stars on an HRD.
However, this in turn requires additional observations such as
multi-wavelength photometry or spectroscopy of a representative sample of the cluster
members.  Second, PMS models must be employed to determine the time varying
mass-luminosity relation.  The accuracy of the derived IMF therefore directly depends on
the accuracy of the adopted PMS models which may be inherently uncertain, particularly
for the youngest clusters ($\tau < 10^6$ yrs)  and lowest mass objects ($m < 0.08$
\msun).  In addition, most PMS models predict bolometric luminosities as a function of mass
and time, and thus bolometric corrections must be used to transform the theoretical
predictions to monochromatic fluxes and magnitudes.

Despite these complexities, Monte Carlo modeling of the infrared luminosity functions of
young clusters (Muench, Lada \& Lada 2000) has demonstrated that {\it the functional
form of an embedded cluster's luminosity function is considerably more sensitive to the
form of the underlying cluster mass function than to any other significant parameter}
(i.e., stellar age distribution, PMS models, etc.).  In fact, despite the significant
differences between the parameters that characterize the various PMS calculations (e.g.,
adopted convection model, opacities, etc.), model luminosity functions were found to be
essentially insensitive to the choice of the PMS mass-to-luminosity relations.  As
discussed below, this reflects the robust nature of PMS luminosity evolution.

There are however other limitations of this technique.  In particular, the observed
luminosity function of a cluster will always contain unrelated foreground and background
field stars along with cluster members.  Such field star contamination can be
straightforwardly corrected for using imaging observations of nearby control fields.
However at the faintest magnitudes, often corresponding to the substellar mass regime,
the field star contamination can be severe and may introduce significant uncertainty in
the faint end of the field star-corrected luminosity function.  


\subsubsection{Individual Stellar Masses from the HR Diagram}

Deriving the IMF of an embedded population via the HRD requires simultaneous
knowledge of both the luminosities and effective temperatures of all the stars in a
cluster so that they can be individually placed on the HRD.  This, in turn,
requires both photometry to determine stellar luminosities and either colors or a
spectrum of each star to determine effective temperature.  For embedded clusters
spectroscopy is the preferred method of obtaining a stellar effective temperature because
the infrared colors of stars are not intrinsically very sensitive to effective
temperature and in addition are significantly altered by extinction and infrared excess
associated with the young stars.  The advantage of this method is that the final product
is the set of individual masses for all stars for which both spectra and photometry
were obtained.  In other words this procedure provides a more detailed determination of
the mass function than the first method.  In addition, this procedure also yields the
ages of the stars and the star formation history of the cluster.  The major disadvantage
of this method is that it requires spectra to be obtained for a complete sample of stars
across the entire spectrum of stellar masses.  As a result, a significant investment of
integration and telescope time is required to obtain a complete sampling of the IMF,
particularly at the faint, low mass end.  As with the first method, field star
contamination, particularly at the lowest masses is a serious limitation.  However, this
limitation can be overcome with the acquisition of spectra for all stars (unrelated
foreground and background stars plus members) within a cluster.  But, at faint
magnitudes, field stars can easily dominate cluster members, sharply decreasing
observing efficiency.  Thus sensitivity limitations inherent in spectroscopic
observations ultimately restrict the application of this technique to a small number of
nearby clusters.

\begin{figure} 
\hskip 0.6in 
\centerline{\psfig{figure=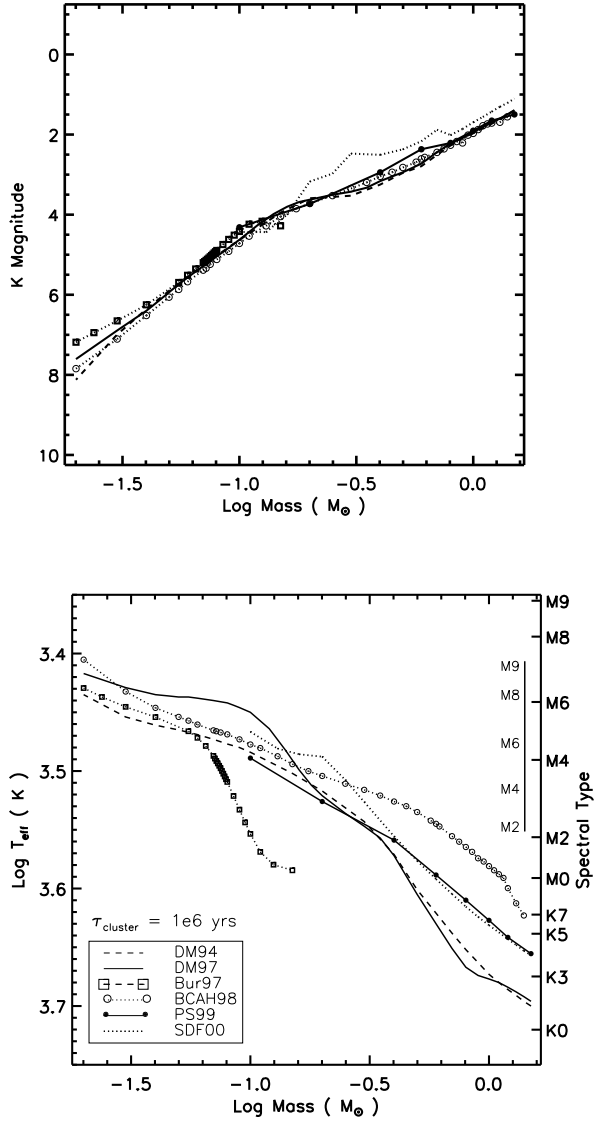,height=60pc}}
\vskip -3.0in 
\caption{Comparison of theoretical predictions for the luminosities and effective
temperatures of million year old PMS stars as a function of mass from a suite of
standard PMS models.  The predicted K magnitudes of such stars (top) are in excellent
agreement over the entire mass spectrum, while the predicted effective temperatures 
(bottom) are
in relatively poor agreement.}  
\label{mlrteff} 
\end{figure}

Similar to the first method, the technique of deriving a PMS star's mass from its
location on the HRD is also fundamentally limited by uncertain knowledge of PMS
evolutionary tracks.  Indeed, the IMF derived by this technique is more sensitive to
uncertainties in PMS models than is the IMF derived by modeling stellar luminosities.
This is because existing PMS models are able to predict the effective temperatures of
PMS stars with considerably less certainty than their luminosities.  This is
illustrated in Figure~\ref{mlrteff} which compares the predictions of standard PMS
models for luminosities and effective temperatures of PMS stars of the same age but
varying mass.  The predicted luminosities are essentially degenerate with respect to the
PMS models used.  Although perhaps surprising at first glance, this result can be
understood by considering the fact that the luminosity of a PMS star is determined by
very basic physics, simply the conversion of gravitational potential energy to radiant
luminosity during Kelvin-Helmholtz contraction.  This primarily depends on the general
physical conditions in the stellar interior (e.g., whether the interior is radiative or
partially or fully convective).  The close agreement of the model predictions reflects
the robust nature of PMS luminosity evolution.  On the other hand, predicting the
effective temperature of such stars, which depends on detailed knowledge of uncertain
characteristics (e.g., opacities) of the stellar atmospheres, is a more difficult
exercise. These same models can predict very different locations for such stars on
the HRD, corresponding to significant differences in the predicted masses 
and mass functions.

\subsubsection{General Limitations}

Other limitations that hinder determinations of the IMFs of embedded clusters via any of
the standard approaches include differential reddening of cluster members, presence of
infrared excess and veiling continuum emission.  These effects need to be accounted for
either in the modeling or by direct correction of the observed photometry and
spectroscopy of individual sources using additional observations.  Embedded clusters are
also at sufficiently large distances that binary systems within them are almost always
unresolved in typical observations.  Consequently, the IMFs derived by these methods do
not include the masses of any unresolved companions.  Binary companions can effect the
derivations of the IMFs in two ways.  First, they can contribute additional flux to the
system luminosity.  However, because the vast majority of binaries are not equal mass
(brightness) systems this contribution is typically small (0.1 - 0.2 magnitudes; e.g.,
see Simon et al.  1995) compared to the typical bin sizes (0.5 mag) used to construct
the infrared luminosity functions.  Second, the presence of unresolved binaries can
result in an underestimate of the numbers of low mass stars in a cluster compared to
that expected for a system of stars in which all binaries are resolved, since companion
stars are not directly observed or counted (Kroupa et al.  1991).  Thus, the IMFs that
are derived are system or primary star IMFs.  Whether or not such a primary star IMF
should be adjusted by adding in the masses of companion stars depends on the question
being considered.  For example, for comparisons with IMFs derived for field stars as
well as open and globular clusters, the primary star IMF is the appropriate IMF to use.
If one desires to exactly weigh the amount of interstellar medium transformed into stars
by the star formation process, then a primary + companion star IMF would be the more
appropriate mass function to consider.  Unfortunately, the IMF of companion stars is not
very well known or constrained by existing observational data and determination of a
total IMF including primary and companion stars is not presently possible.  Finally,
since the IMF is a statistical property of an ensemble of stars, it can only be
meaningfully derived over a mass interval which is statistically well sampled by
observations.  The richness of the observed cluster thus sets a basic limit on the level
of uncertainty in any derived IMF.

\subsection{The IMF of the Trapezium Cluster from OB Stars to Brown Dwarfs}

The Trapezium cluster in Orion is the best studied of all embedded clusters.  First
identified by Trumpler (1931), the Trapezium cluster is a rich cluster of faint (mostly
PMS) stars embedded within the Great Orion Nebula with an age of about 10$^6$ yrs
(Prosser et al. 1994; Hillenbrand 1997).  The
cluster is approximately 0.3 - 0.4 pc in diameter (e.g., Lada et al.  2000) and contains
approximately 700 stars (Hillenbrand \& Carpenter 2000; Muench et al 2002).  It is
thought to be the highly concentrated core of the more extended Orion Nebula Cluster
(ONC) which contains nearly 2000 stars spread over a region roughly 4 pc in
extent (e.g., Hillenbrand \& Hartmann 1998).  At its center is the famous Trapezium, a
close grouping of four OB stars which excite the nebula.  It is a superb target for IMF
studies because of its youth, richness, compactness, location in front of and partially
within an opaque molecular cloud, and its proximity to the sun ($\sim$ 450 pc).  These
factors combine to enable a statistically significant sampling of the IMF from OB stars
to substellar objects near the deuterium burning limit ( 0.01 \msun) with minimal field
star contamination.  Indeed, this cluster is particularly well suited for investigating
the substellar portion of the IMF and determining the initial distribution of masses for
freely floating brown dwarfs.  Deep infrared surveys of this cluster have been performed
using the HST (Luhman et al 2000), the Keck Telescope (Hillenbrand \& Carpenter 2000),
UKIRT (Lucas \& Roche 2000) and the NTT (Muench et al 2002) and have produced infrared
luminosity functions and mass functions which sample well into the substellar mass
range.

\begin{figure} 
\vskip 6.0in

\epsfxsize=7.5in 


\caption{Optical (top) and deep {\it JHK} infrared (bottom) images of
the Trapezium cluster in Orion obtained with the NASA HST and the ESO VLT, respectively.  
The infrared observations taken from Muench et al. 2002}  
\label{trapntt}
\end{figure}

\begin{figure} 
\epsfxsize=5.0in 
\centerline{\epsfbox{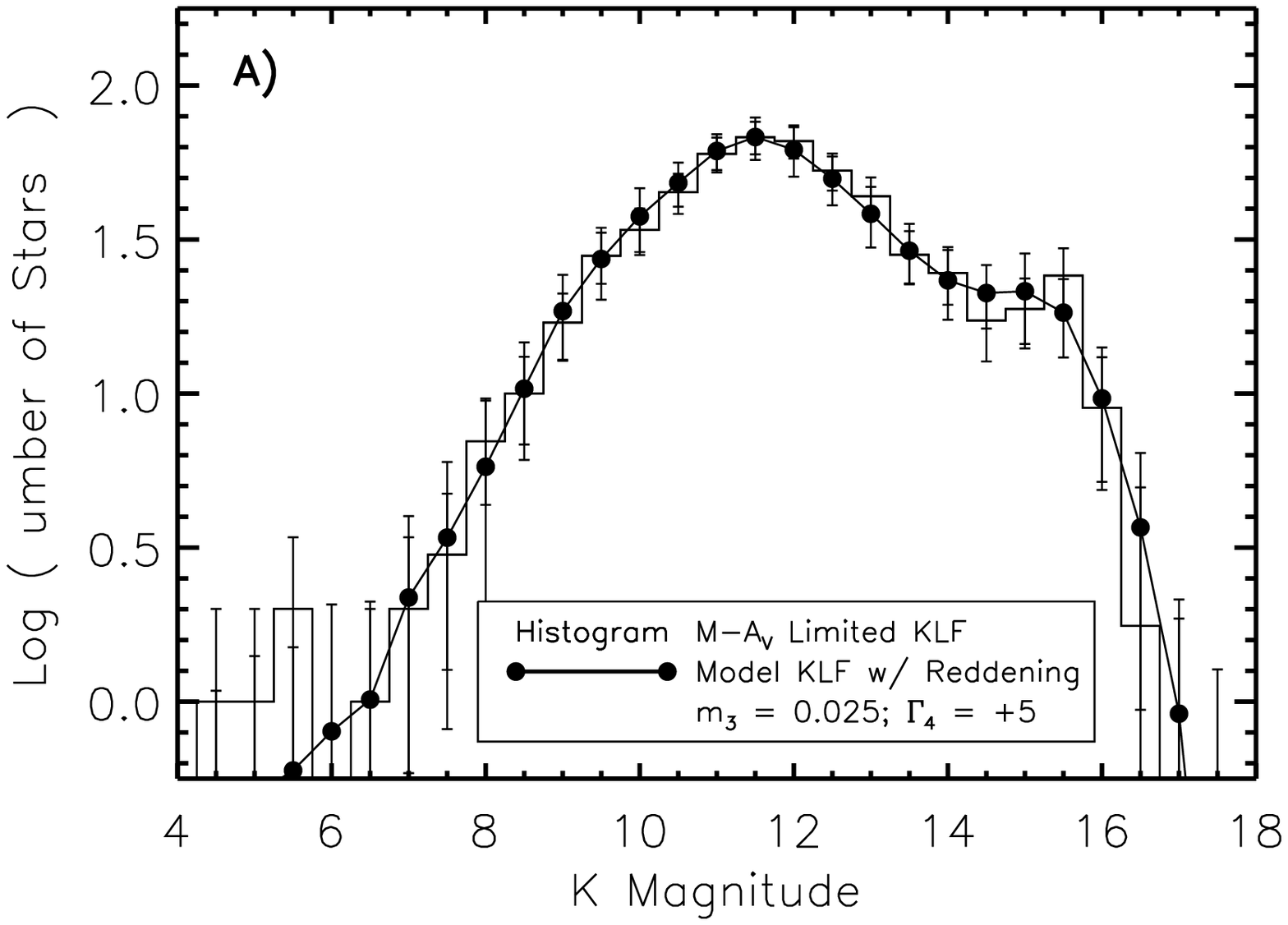}} 
\caption{The K-band (2.2$\mu$m) monochromatic
luminosity function (histogram) of the Trapezium cluster constructed from the deep
infrared imaging observations shown in Figure~\ref{trapntt}.  This is a background
corrected, extinction-limited KLF complete to a cloud depth of A$_V$ = 17 magnitudes for
million year old stars with masses greater than $\sim$ 0.015 \msun.  Also plotted is the
model KLF (line) corresponding to the underlying IMF (Figure~\ref{trapimf}) whose KLF
best fit the data.  Taken from Muench et al.  (2002)} 
\label{trapklf} 
\end{figure}

Figure~\ref{trapntt} shows a three-color infrared image of the cluster resulting from
the NTT survey.  Muench et al.  (2002) used this data along with observations of the
same region obtained with a 1.2 meter telescope to recover the brighter stars typically
saturated in deep exposures with the larger telescopes and they produced a complete
sampling of the K-band (2.2 $\mu$m) luminosity function (KLF) of this cluster spanning
the mass range from OB stars to substellar objects near the deuterium burning limit.
Figure~\ref{trapklf} shows the field-star corrected, complete, extinction-limited, KLF
derived from the Muench et al.  study.  It counts all stars within a cloud depth of
17 magnitudes of visual extinction with luminosities corresponding to million year old
objects with masses $\sim$ 0.010 -- 0.015 \msun and greater and is representative of the
infrared luminosity functions obtained in all similarly sensitive investigations of this
cluster.  In particular, the KLF is found to rise steadily from the brightest stars to
around m$_K$ $\sim$ 11 -12 mag where it flattens before clearly falling again to fainter
magnitudes.  A clear secondary peak is present at approximately 15th magnitude, which is
well into the brown dwarf luminosity range.  At lower luminosities the KLF rapidly drops
off.

Muench et al.  derived the IMF of the Trapezium cluster by using a suite of Monte Carlo
calculations to model the cluster's KLF.  The observed shape of a cluster luminosity
function depends on three parameters:  the ages of the cluster stars, the cluster
mass-luminosity relation, and the underlying IMF (i.e., Equation~\ref{eq1}).  With the
assumptions of a fixed age distribution, derived from the spectroscopic study of the
cluster by Hillenbrand (1997), a composite theoretical mass-luminosity relation adopted
from published PMS calculations (i.e., Bernaconi 1996; Burrows et al.  1997; D'Antona \&
Mazzitelli 1997; Schaller 1992), and an empirical set of bolometric corrections, Muench
et al.  varied the functional form of the underlying IMF to construct a series of
synthetic KLFs.  These synthetic KLFs were then compared to the observed Trapezium KLF in
a Chi-Squared minimization procedure to produce a best-fit IMF.  As part of the modeling
procedure, the synthetic KLFs were statistically corrected for both variable extinction
and infrared excess using Monte Carlo probability functions for these quantities derived
directly from multi-color (JHK) observations of the cluster.

\begin{figure} 
\epsfxsize=5.8in 
\centerline{\epsfbox{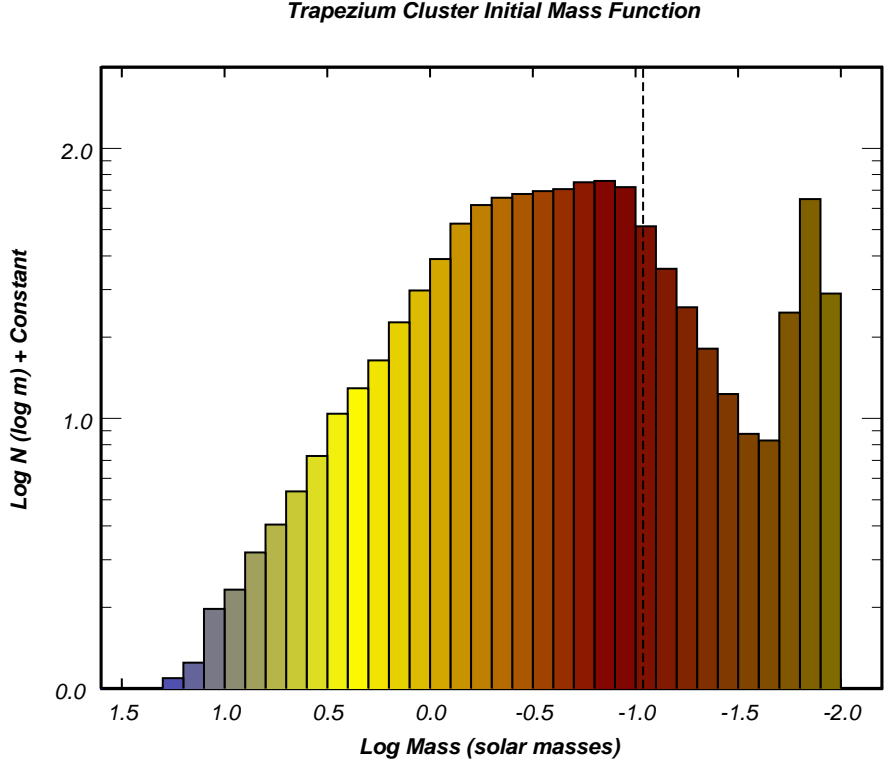}}
\caption{The IMF derived for the Trapezium
cluster from Monte Carlo modelling of its luminosity function (Muench et al.  2002).
This plot displays the binned mass function of the synthetic cluster whose luminosity
function was found to best fit the observed KLF of the Trapezium cluster.  (See
Figure~\ref{trapklf}) A vertical dashed line marks the approximate location of the
hydrogen burning limit (HBL).  The derived IMF displays a broad peak between 0.1 - 0.6
\msun and extends deep into the substellar mass regime.  The secondary peak is located
near 0.015 \msun or 15 $M_J$.  It corresponds to the bump in the KLF at K $\sim$ 15.5
magnitudes seen in Figure~\ref{trapklf} and may be an artifact of the adopted substellar
MLR.}  
\label{trapimf} 
\end{figure}

The best-fit synthetic KLF is plotted in Figure~\ref{trapklf}.  The corresponding
underlying mass function is displayed in Figure~\ref{trapimf} in the form of a histogram
of binned masses of the stars in the best-fit synthetic cluster.  This model mass
function represents the IMF of the young Trapezium cluster.  The main characteristics of
this IMF are:  1) the sharp power-law rise of the IMF from about 10 \msun (OB stars) to
0.6 \msun (dwarf stars) with a slope (i.e., $\beta$ = --1.2) similar to that of Salpeter
(1955), 2) the break from the single power-law rise at 0.6 \msun followed by a flattening
and slow rise reaching a peak at about 0.1 \msunp, near the hydrogen burning
limit, 3) the immediate steep decline into the substellar or brown dwarf
regime and finally 4) the prominent secondary peak near 0.015 \msun or 15 $M_J$ (Jupiter
masses) followed by a very rapid decline to lower masses beyond the deuterium burning
limit (at $\sim$ 10 $M_J$).

The most significant characteristic of this IMF is the broad peak, extending roughly from
0.6 to 0.1 \msunp.  {\it This structure clearly demonstrates that there is a
characteristic mass produced by the star formation process in Orion.}  That is, the
typical outcome of the star formation process in this cluster is a star with a mass
between 0.1 and 0.6 \msunp.  The process produces relatively few high mass stars and
relatively few substellar objects.  {\it Indeed, no more than $\sim$ 22\% of all the
objects formed in the cluster are freely floating brown dwarfs}.  The overall continuity
of the IMF from OB stars to low mass stars and across the hydrogen burning limit strongly
suggests that the star formation process has no knowledge of the physics of hydrogen burning.
Substellar objects are produced naturally as part of the same physical process that
produces OB stars (see also Najita, Tiede \& Carr 2000; Muench et al.  2001).

In this respect the secondary peak at 0.015 \msun is intriguing.  The existence of such a
peak may imply a secondary formation mechanism for the lower mass brown dwarfs, similar
to suggestions recently advanced by Reipurth and Clarke (2000) and thus is potentially
very important.  However, the significance that should be attached to this feature
depends on the accuracy of the adopted mass-luminosity relations for substellar objects
used in the modelling.  These MLRs may be considerably more uncertain than those of PMS
stars.  Indeed, observations of an apparent deficit of stars in the M6-M8 range of
spectral types in a number of open clusters are suggestive of the existence of a
previously unknown opacity feature in the MLR for such cool stars (Dobbie et al.  2002).
The presence of such a feature could produce a peak in the luminosity function and a
corresponding artificial peak in the derived mass function if not included in the 
theoretical MLRs
(e.g., Kroupa Tout \& Gilmore 1990, 1993).  Given that this spectral type range
corresponds to the temperature range predicted for young low mass brown dwarfs, it is
quite plausible that the secondary feature in the derived IMF is artificial and does not
represent a true feature in the underlying IMF.  Clearly more data, both observational
and theoretical is needed to assess the reality and significance of this intriguing
feature.

\begin{figure} 
\epsfxsize=5.0in 
\centerline{\epsfbox{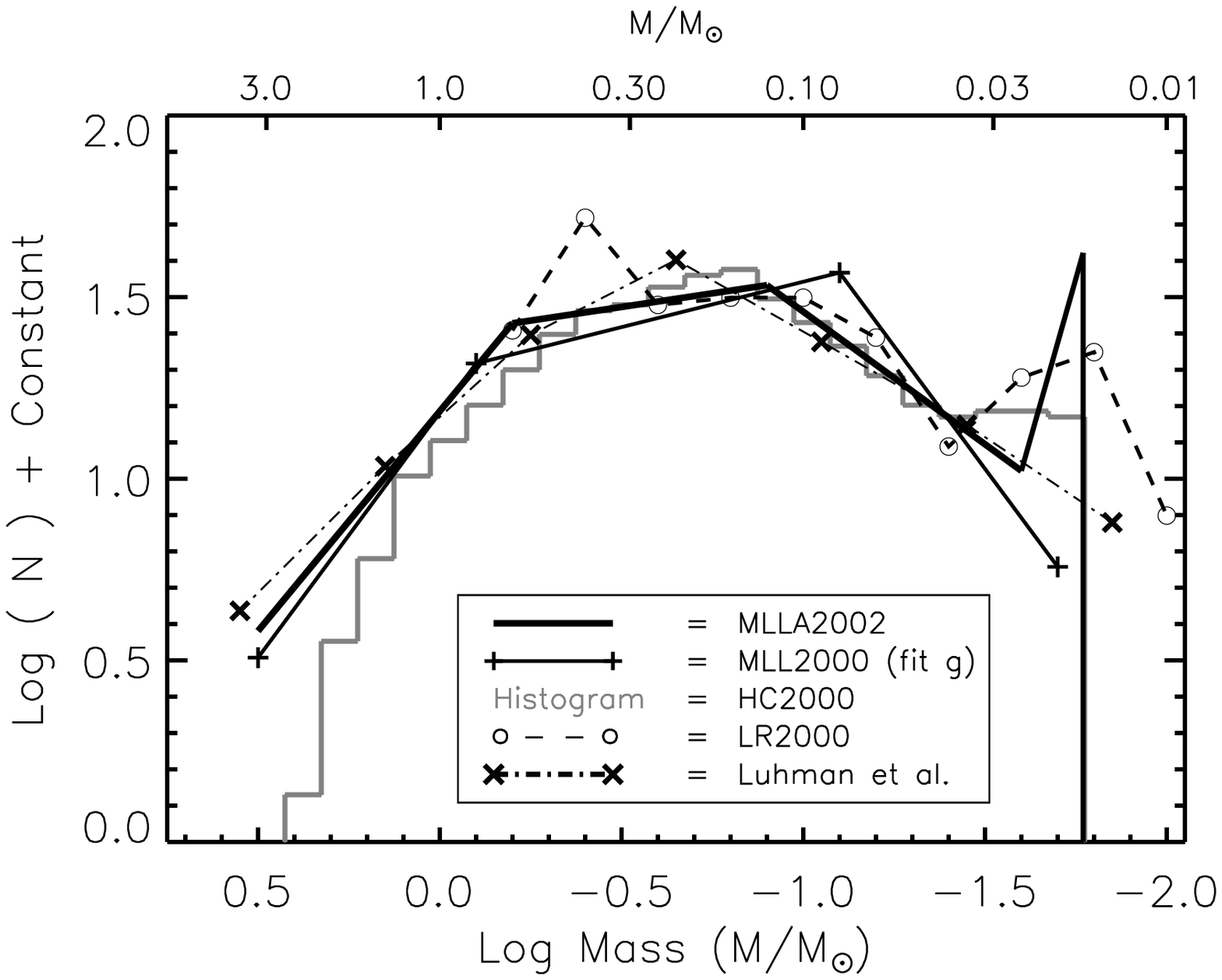}} 
\caption{Comparison of IMFs derived from infrared
observations reported in recent studies in the literature.  These IMFs, derived from
independent observations using a variety of techniques, are in very good agreement
concerning the basic functional form of this cluster's IMF.  In particular, all exhibit
a broad peak extending from roughly 0.6 - 0.1 \msun followed by a clear decline in the
substellar regime.  This suggests a characteristic mass is produced by the star
formation process.  } 
\label{trapcomp} 
\end{figure}

Figure~\ref{trapcomp} shows a comparison of Trapezium IMFs recently derived from a number
of different deep infrared imaging surveys using a variety of methods (Lucas \& Roche
2000; Hillenbrand and Carpenter 2000; Muench et al.  2000; Luhman et al.  2000; Muench et
al.  2003).  The general agreement of the derived IMFs is impressive for $m_* >$ 0.015
\msunp.  The fundamental features (1-3) described above are evident in all the IMFs.  In
the region of the secondary peak (4), the agreement is less impressive, likely reflecting
the inherent uncertainties in the modeling at the lowest substellar masses.
Nonetheless, there is general agreement that the IMF turns over and falls off below the
hydrogen burning limit and into the substellar mass regime.  However, the precise
details, such as the steepness of the falloff and the amplitude of the secondary peak,
remain somewhat uncertain.  Other differences in details between the various IMFs likely
result from the uncertainties inherent in the different techniques used in the IMF
determinations and provide some measure of the overall uncertainty in our present ability
to measure the exact form of the IMF in this cluster.  Clearly, however, these infrared
studies of the Trapezium cluster have established the fundamental properties of its IMF.

The derived IMF of the Trapezuim cluster spans a significantly greater range of mass than
any previous IMF determination whether for field stars or other clusters (e.g., Kroupa
2002).  Its statistically meaningful extension to substellar masses and the clear
demonstration of a turnover near the HBL represents an important advance in IMF studies.

For masses in excess of the HBL the IMF for the Trapezium is in good agreement with the
most recent determinations for field stars (Kroupa 2002).  This is to some extent both
remarkable and surprising since the field star IMF is averaged over billions of years of
galactic history, assuming a constant star formation rate, over the age of the Galaxy,
and over stars originating from very different locations of galactic space.  The
Trapezium cluster, on the other hand, was formed within the last million years in a
region considerably less than a parsec in extent.  Moreover, there is evidence that this
region is not yet finished producing stars as significant star formation appears to be
continuing in the molecular cloud behind the cluster (Lada et al 2000).  Taken at face
value this agreement suggests that the IMF and the star formation process that produces
it is very robust, at least for stellar mass objects.

\subsection{Comparison With Other Embedded Clusters:  A Universal IMF?}

Although few other embedded clusters have been as completely studied as the Trapezium,
the luminosity and mass functions of a number of such clusters have been investigated in
various levels of detail.  These include such clusters as IC348 (Lada \& Lada 1995;
Luhman et al.  1998; Najita, Tiede \& Carr 2000; Muench et al.  2003), NGC 1333 (Aspin,
Sandell \& Russell 1994; Lada,
Alves \& Lada 1996), NGC 2264 (Lada, Young \& Greene 1993), NGC 2024 (Lada et al.  1991b;
Comeron et al.  1996; Meyer 1996), Rho Ophiuchi (Lada and Wilking 1984; Comeron et al.
1993; Green and Meyer 1995; Bontemps et al. 2001), Serpens (Eiroa \& Casali 1992; Giovannetti et al.  1998),
M17 (Lada et al.  1991a), W3 (Megeath et al.  1996), and NGC 3603 (Eisenhauer et al.
1998; Brandl et al.  1999; Nurnberger \& Petr-Gotzens 2002).  Due varying distances,
sizes, sensitivities, methodologies, etc., these KLFs and corresponding IMFs were not
uniformly sampled nor investigated using a common systematic approach.  As a result only
limited conclusions can be drawn from comparison of all these results with each other and
with the IMF derived for the Trapezium cluster.  On the other hand, when homogeneous data
is analyzed with similar methodology more meaningful comparisons of embedded cluster IMFs
are possible (e.g., Lada et al.  1996; Luhman et al.  2000; Muench et al.  2002).

\begin{figure}
\epsfxsize=5.0in
\centerline{\epsfbox{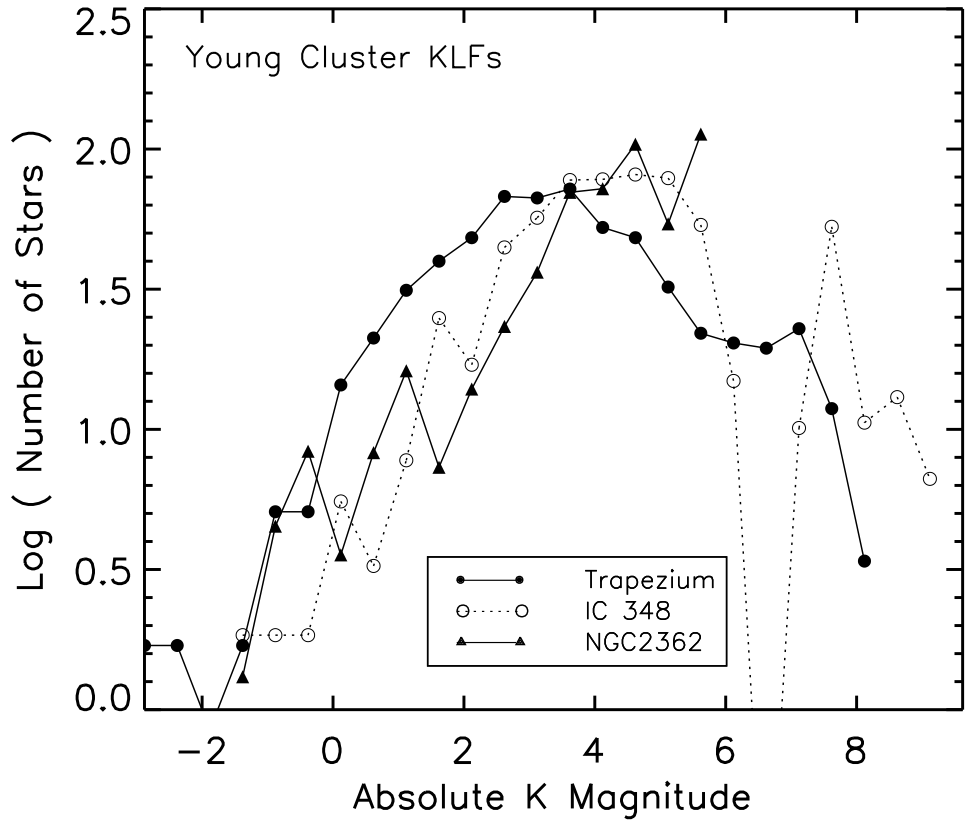}} 
\caption{ Comparison of the observed KLFs (adjusted to the same distance) of 
three young clusters of 
differing age: Trapezium ($10^6$ yrs), IC 348 ($3 \times 10^6$ yrs) and
NGC 2362 ($5 \times 10^6$ yrs). A significant trend toward lower luminosity
with age is apparent for these KLFs, similar to that predicted by evolutionary models.
see Figure~\ref{klfevo}. }
\label{klfcomp} 
\end{figure}

\begin{figure}
\epsfxsize=5.0in
\centerline{\epsfbox{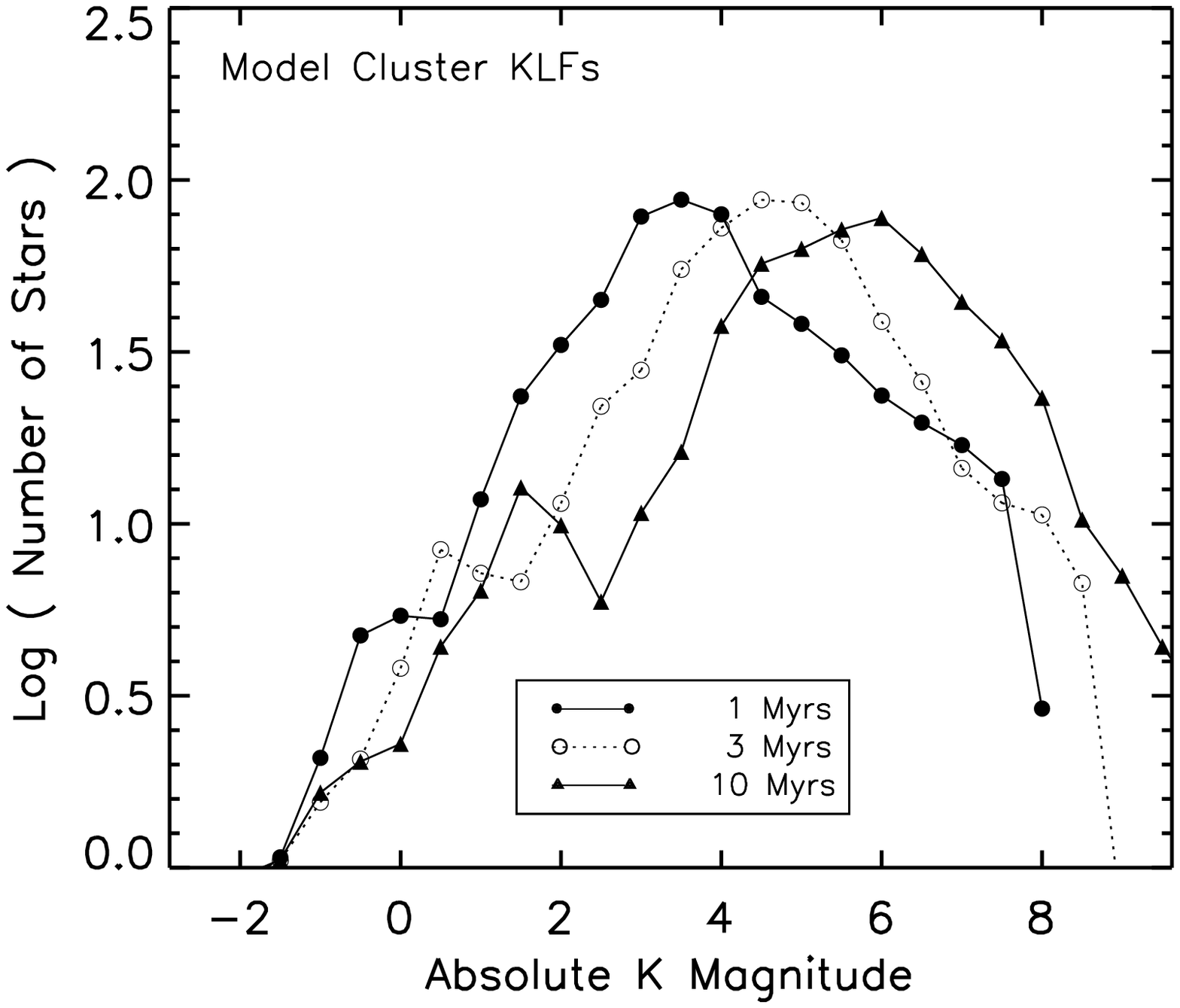}} 
\caption{Model luminosity functions (KLFs) of synthetic clusters of differing age
but with the same underlying IMF (that of the Trapezium cluster).
Young clusters experience systematic luminosity evolution,
gradually becoming fainter with time as PMS stars within them approach the 
main sequence. Figure prepared by August Muench. }
\label{klfevo} 
\end{figure}

The first conclusion that can be drawn from such studies is that the KLF for embedded
clusters is not a universal function and statistically significantly variations are
present in observed clusters (e.g., Lada, Alves \& Lada 1996).  This is illustrated in
Figure~\ref{klfcomp} which displays the KLFs of the Trapezium, IC 348 and NGC 2362 clusters.
Such variations in the cluster luminosity functions are not unexpected for embedded
clusters which consist mostly of PMS stars.  Even if such clusters were characterized by
a universal IMF, they would experience significant luminosity evolution as their PMS
population evolved and collectively approached the main sequence.  This luminosity
evolution would be particularly rapid during the first 5 Myrs of a cluster's existence
when the luminosities of its PMS stars experience their most rapid declines.  For
example, Figure~\ref{klfevo} shows the expected KLFs for three different aged synthetic
clusters with identical IMFs.  These KLFs were calculated from Monte Carlo simulations
by Muench et al.  (2000).  The systematic evolution of the KLFs to lower brightness is
clearly evident and certainly would be significant enough to be observable.  Indeed,
modeling of the KLFs of clusters such as IC 348, NGC 2362 and the Trapezium, indicates
that observed differences in their KLFs  can be explained by the expected luminosity
evolution in clusters that have very similar underlying mass functions
(Lada \& Lada 1995; Muench et al.  2002; Alves et al.  2003).
This can also be inferred empirically, without resorting to modeling.  Comparison of the
KLFs of young clusters of known age shows that clusters of similar age display KLFs of
very similar shape while clusters of differing age show the greatest variation of KLF
forms (Lada, Alves \& Lada 1996; Alves et al.  2003).

\begin{figure}[t] 
\vskip 2.0in
\centerline{\psfig{figure=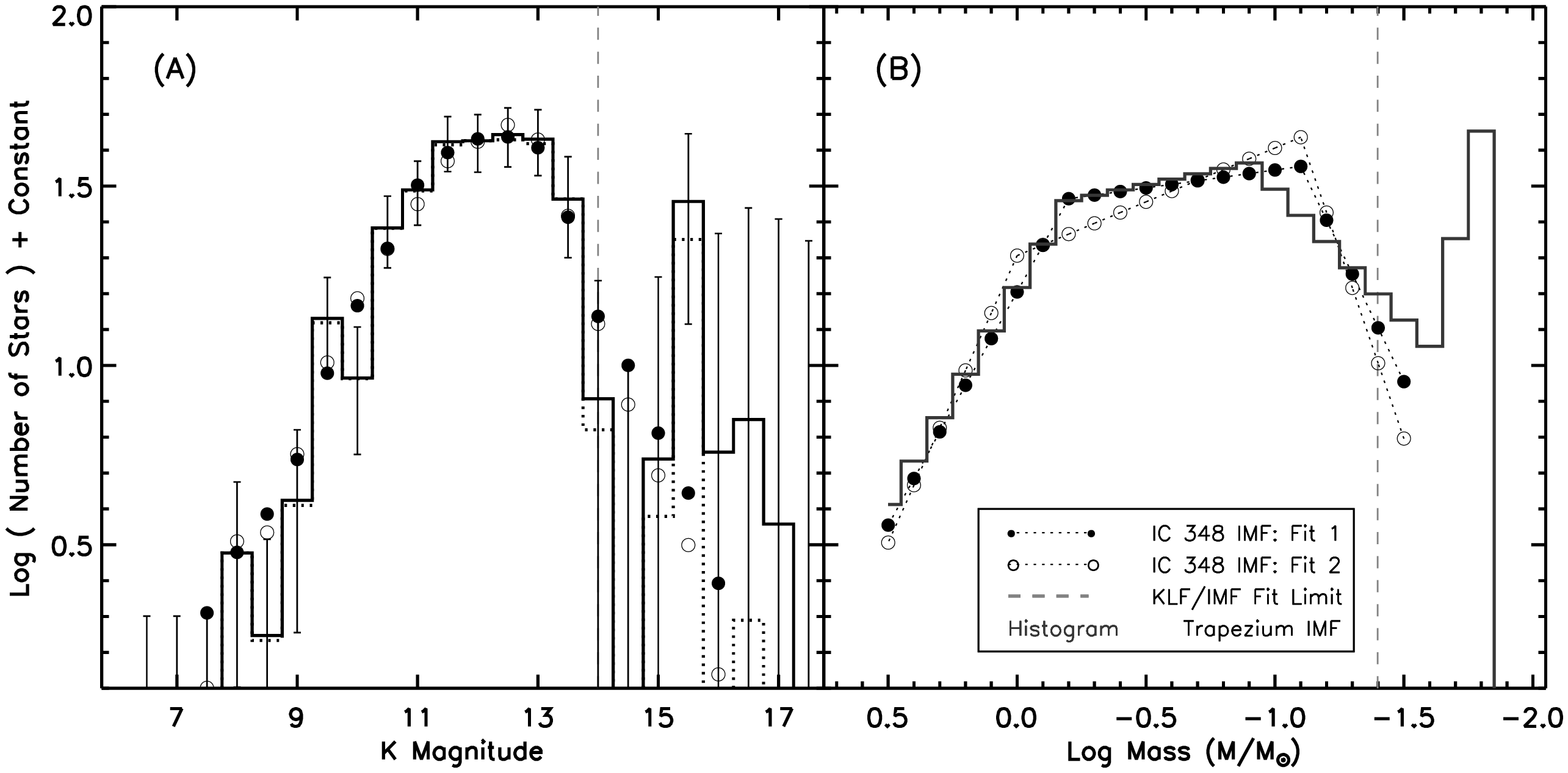,width=5in,angle=90}}

\caption{The
KLF and derived IMF for IC 348.  The left panel shows the observed background corrected
KLF (histogram) along with the best fit synthetic KLFs (filled and open circles)
corresponding to the underlying model IMFs (filled and open circles) shown in the right
hand panel.  Also plotted is the IMF derived for the Trapezuim cluster.  From Muench et
al.  (2003).}  
\label{ic348imfklf} 
\end{figure}

To illustrate this further we display in Figure~\ref{ic348imfklf} the background
corrected KLF and corresponding IMF derived for the IC 348 cluster from Monte Carlo
modeling (Muench et al.  2003).  IC 348 is the best studied embedded cluster after the
Trapezium.  Although it is not as rich and suffers appreciably more background
contamination, this cluster is closer to the solar system, older ($\tau \sim $ 2-3
Myr.)  and more evolved than the Trapezium cluster (Lada \& Lada 1995; Herbig 1998).
This places additional constraints for IMF modeling making the cluster a good candidate
for comparison with the Trapezium.  As seen in the figure, the IMF for this cluster
derived from KLF modeling compares very well with that of the Trapezium over the entire
range of mass ($m_* > \sim 15 M_J$) over which it was determined.  In particular, the
IMF displays a broad peak between 0.1 - 0.5 \msun with a clear turn down near the HBL.
This indicates that there is a characteristic mass produced by the star formation
process in IC 348 and this mass is essentially the same as that suggested for star
formation in the Trapezium.  Moreover, the size of the substellar population of IC 348
is relatively well determined and constitutes only about $\sim$ 25 \% of the cluster
membership, again similar to the Trapezium.  Because of poor statistics, Muench et al.
did not fit the KLF in the luminosity range that corresponds to the location of the
secondary peak in the substellar IMF of the Trapezium. However examination of the
IC 348 KLF shows a marginally significant bump at the luminosity corresponding to that
mass.  The relatively poor statistics in the IC 348 KLF at these faint magnitudes
is due to a combination of the fact that IC 348 is not as
rich as the Trapezium and suffers more contamination from background stars since it is
not foreground to a similarly opaque background molecular cloud.

Using independent observations and different methodology to extract IMFs from the
observations, Luhman et al.  (2000) derived IMFs for the Trapezium, IC 348 and Ophiuchi
clusters and also found good agreement in the shapes of the IMFs of these clusters above
the hydrogen burning limit.  These comparisons again suggest a relatively robust or
universal IMF.  Another indication of the robustness of the IMF is found in the
observations of NGC 3603, perhaps the most luminous and richest young cluster in the
Galaxy.  This distant ($\sim$ 7 Kpc), massive and very dense OB star cluster has been
shown to contain low mass (0.5 - 1 \msun) stars in roughly the proportion predicted for
a normal IMF (Nurnberger \& Petr-Gotzens 2002).  This is interesting given the
significant differences in cluster richness (and central densities) which likely
corresponds to differences in formation environment and internal star formation rates.
Indeed, it is entirely possible that the formation process for the Ophiuchi, NGC 1333
and even perhaps the Trapezium clusters is far from being completed.  Yet, it seems that
once even a hundred or so stars are formed the fundamental form of the IMF is already
determined.  Differences in the underlying IMFs of these clusters may exist, but they
are not large enough to be detectable given the existing uncertainties in the
measurements, which are likely dominated by the small sizes of the existing samples.

It is interesting as discussed earlier that a consensus appears to be building that IMFs
for open clusters can also be described by a universal form similar to the field star IMF
(Massey 1998; Kroupa 2002).  Here we note that recent studies of the Pleiades (Bouvier et
al.  1998; Adams et al.  2001) and $\alpha$ Per (Barrado y Navascues et al.  2002)
clusters have extended the IMFs of these older (10$^8$ Myr.)  clusters to the substellar
regime and thus an initial comparison can be made with the embedded clusters.  These open
cluster IMFs display the same 4 fundamental characteristics described above for the
Trapezium cluster, including a secondary peak in the low mass portion of the substellar
IMF!  The IMFs do, however, differ in the location of the substellar peak.  In the two
older open clusters this peak is found between 40-50 $M_J$, a significantly higher mass
than suggested ($\sim$ 15 $M_J$) by observations of the Trapezium and IC 348 clusters.
This perhaps indicates that the feature is indeed the result of a hitherto
unknown feature in the brown dwarf MLR (Dobbie et al.  2002) as discussed earlier.

Strong similarities have emerged in the IMFs derived for field stars, open clusters and
embedded clusters suggesting a robust and universal IMF.  However, there is some evidence
that this universal IMF may not characterize all star formation events.  For example, HST
observations of the extragalactic star burst cluster R136 (30 Dor) in the Large
Magellanic Cloud suggest that the IMF of this O star rich cluster flattens and departs
from a Salpeter-like power-law rise at a mass of about 2 \msun (Sirianni et al.  2000).
This is higher than the mass ($\sim$ 0.6 \msun) of the corresponding inflection point in
the Trapezium, open cluster and field star IMFs and suggests a relative deficit of lower
mass stars in this rich O cluster.  However, it is likely that the numbers of solar mass
stars in R136 has been underestimated due to the severe crowding in the cluster center.
It is still quite possible that even in this cluster, which contains 1000 O stars, the
underlying IMF is characterized by the same universal form as that derived for the
Trapezium.  A more significant indication of a departure from a universal IMF may be
present in recent observations of the nearby Taurus clouds.  Luhman (2000) has found a
significant (factor of 2) deficit of substellar mass objects in this region relative to
the embedded population in the Trapezium and IC 348 clusters.  The embedded population of
the Taurus clouds is an embedded T association consisting of isolated stars and small
loose groupings of stars formed over a relatively large area.  These conditions are
decidedly different than those which characterize embedded cluster formation.  Since for
Taurus the IMF above the HBL appears to be similar to that of clusters and the field
(e.g., Kenyon \& Hartmann 1995), the finding of a deficit of brown dwarf stars may
indicate that the substellar IMF is less robust than the stellar IMF and thus may be a
sensitive function of formation environment and/or initial conditions.  However, more
observations would be necessary to test the significance of this possibility.

\section{LABORATORIES FOR STAR AND PLANET FORMATION}

Nearly half a century ago, Walker's (1956) observations of the partially embedded cluster
NGC 2264 showed that its late type (i.e, F and later) stars were characterized by subgiant
luminosities which placed them well above the main sequence on the HRD.  Thus, these
observations empirically established the pre-main sequence, pre-hydrogen burning, nature of
young low mass stars and provided the critical data needed to test and constrain the theory
of PMS evolution (e.g., Hayashi 1966).  Thirty years later infrared observations of the
embedded cluster in Ophiuchus enabled the first systematic classification of infrared
protostars and young stellar objects based on emergent stellar energy distributions
(Wilking \& Lada 1983; Lada \& Wilking 1984, Lada 1987).  Such observations were very
influential in constructing the early framework for a theoretical understanding low mass
star formation (e.g., Shu, Adams \& Lizano 1987).  Today, embedded clusters continue to
play an important role for the development and testing of theories dealing with the
formation and early evolution of both stars and planetary systems.

\subsection{Protostars and Outflows}

Embedded clusters contain young stellar objects in various evolutionary states, including
deeply buried protostars with their infalling envelopes and associated bipolar outflows and
more exposed PMS stars surrounded by protoplanetary accretion disks.  Clusters are
particularly useful for statistical studies of the evolution of such objects.  For example,
infrared photometric and spectroscopic observations have been used in comparative studies
to investigate the physical and evolutionary natures of embedded populations in clusters.
Observational investigations of the embedded Ophiuchi cluster led to the identification of
four classes of young stellar objects corresponding to four phases of early stellar
evolution (Lada \& Wilking 1984; Adams, Lada \& Shu 1987; Andre, Ward-Thompson \& Barsony
1993).  The ratios of the numbers of objects in the various stages coupled with cluster
ages have led to estimates for the lifetimes of the various states of early stellar
evolution.  For example, the rarity of protostars in even the youngest clusters suggests
that the protostellar phases are relatively short, $\sim$ 10$^4$ yrs to $\sim$
10$^5$ yrs  (e.g., Wilking, Lada \& Young 1989; Greene et al 1994, Andre \&
Montmerle 1994).  Coupled with observation of source luminosities and estimates of the
masses of the underlying stars, these timescales yield mass accretion rates for
protostellar evolution which constrain theoretical predictions (Shu, Adams \& Lizano 1987).

More recently, spectroscopic surveys of embedded clusters have been employed to
systematically investigate the more detailed physical properties of embedded populations.
Low resolution infrared spectroscopic surveys of the embedded population in the Rho
Ophiuchi cluster revealed significant differences between protostellar objects and PMS
stars indicating that protostellar photospheres were heavily veiled.  The large measured
veilings are likely the result of excess emission originating in infalling envelopes and
surrounding accretion disks (Casali \& Matthews 1992; Casali \& Eiroa 1996; Greene \& Lada
1996).  Deep, high resolution spectral observations of the embedded populations of the
Ophiuchi cluster further found the rotation rates of classical protostars (class I sources)
to be systematically greater than those of disk bearing PMS stars (class II sources) in the
cluster suggesting that rotation rates in young stellar objects may be modulated by their
accretion rates (Greene \& Lada 1997, 2002).

The demographics of the protostellar populations in embedded clusters are not well known
and poorly constrained by existing observations.  Although generally rare, protostars are
most abundant in the youngest embedded clusters where as many as 15-20\% of the members can
be protostellar in nature.  Examples of embedded clusters that are protostar rich include
NGC 1333, Serpens and Rho Ophiuchi.  Likely no more than 10$^6$ yrs old, these clusters
represent the least evolved embedded clusters known.  Indeed, both NGC 1333 and the Serpens
cluster each contain relatively large populations of class 0 sources, the most deeply
embedded and least evolved protostellar class identified (Sandell \& Knee 2001; Hurt \&
Barsony 1996).  These two clusters are also rich in bipolar outflows and both appear to be
undergoing recent bursts in outflow activity (Davis et al.  1999; Bally, Devine \& Reipurth
1996; Knee \& Sandell 2000).  Both NGC 1333 and Serpens are experiencing vigorous star
formation and are likely true protoclusters, still largely in the process of being
constructed from molecular gas and dust.  This is in contrast to the situation for older
embedded clusters such as IC 348, which, at an age of $\sim$ 2-3 Myr., appears to have few
protostellar sources and only one outflow (Lada \& Lada 1995; McCaughrean, Rayner, \&
Zinnecker 1994).  Lada et al.  (1996) have argued that NGC 1333 and IC 348 have similar
present day rates of star formation and suggested that NGC 1333 would evolve into a cluster
similar to IC 348 if star formation were to continue in NGC 1333 at the same rate for
another 2-3 Myr.  However, once a cluster's age exceeds the timescale ($\sim$ 10$^5$ yrs)
for protostellar evolution, the older and more evolved the cluster becomes, the smaller the
fraction of protostars and outflows it will contain (e.g., Fletcher \& Stahler 1994b).
Since most embedded clusters have ages between 1--3 Myrs, their membership is
typically dominated by PMS stars.

\subsection{Circumstellar-Protoplanetary Disks}

PMS stars come in two varieties:  those with circumstellar disks (Class II) and those
without such disks (Class III).  The frequency of disks within a cluster is directly
related to the physical processes of disk formation and evolution.  Since circumstellar
disks can be the progenitors of planetary systems, knowledge of the cluster disk fraction
(CDF) and how it evolves with time also have important consequences for understanding the
origin of planetary systems.  Since most stars probably formed in embedded clusters, the
measurement of the CDF in the youngest embedded clusters produces a determination of the
initial disk frequency (IDF) which in turn directly measures the probability of disk
formation around newly formed stars.  Coupled with knowledge of the probability of planet
formation in circumstellar disks, the IDF can provide an estimate and indirect census of
extrasolar planetary systems in our galaxy.  The variation of the CDF with cluster age
sets the timescale for disk evolution and thus the duration or lifetime of the
circumstellar disk (and planet building) phase of early stellar evolution.  This
therefore provides a critical constraint for determining the probability of planet
formation in circumstellar disks and so therefore also directly bears on the question of
the ubiquity of extrasolar planetary systems.  Do stars of all masses form with
circumstellar disks?  Is the likelihood of forming planetary systems from disks dependent
on the mass of the central star, or on the environment in which the star formed, or both?
These and similar questions concerning the origins of planetary systems may be best
addressed by observations of embedded clusters.

\begin{figure}
\epsfxsize=5.0in
\centerline{\epsfbox{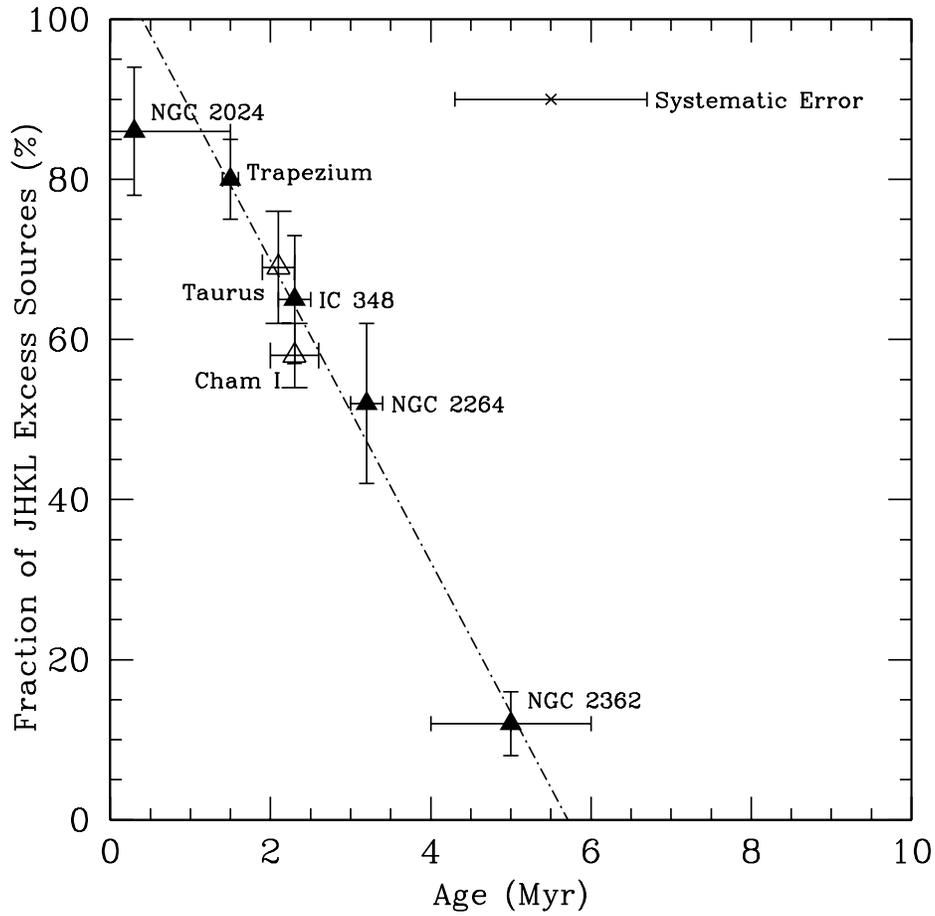}}
\caption{Disk Fraction as a function of cluster age for a sample of young clusters
with consistently determined mean ages. The disk fraction is initially very high,
but then rapidly drops with cluster age suggesting maximum disk lifetimes of 
less than 6 Myrs in young clusters (Haisch et al. 2001a)}
\label{diskfreq} 
\end{figure}

The CDF and its dependence on stellar mass and cluster age, in principle, can be directly
measured by obtaining the infrared spectral energy distributions of all or a
representative fraction of a cluster population.  This is because stars with
circumstellar disks emit excess infrared emission which displays a clear and specific
spectral signature in the star's optical-infrared energy distribution (e.g., Lynden-Bell
\& Pringle 1974; Adams, Lada \& Shu 1987).  In practice, it is prohibitive to obtain
complete (i.e., 1-1000 $\mu$~m ) infrared spectral energy distributions of a cluster
population since to do so requires multi-wavelength ground-based and space-based
observations.  However, an excess from a disk can be measured at any infrared wavelength
sufficiently longward of the peak of the underlying stellar energy distribution of the
central star.  Although, the longer the wavelength the more unambiguous the infrared
excess, observations at wavelengths as short as 2 $\mu$~m (K-band) can detect infrared
excesses in the majority of disk bearing stars.  Indeed, essentially all stars with
circumstellar disks containing 10$^{-9}$ \msun of hot dust or more can be detected using
ground-based L-band (3.4 $\mu$m) observations (Lada et al. 2000; Wood et al.  2002).

Near-infrared JHK imaging surveys of numerous embedded clusters have suggested that the
IDF is relatively high ($>$ 50\%; e.g., Stauffer et al.  1994; Lada, Alves \& Lada
1996; Carpenter et al 1997).  Indeed, longer wavelength (i.e, 3-10 $\mu$~m) surveys,
which are more sensitive to disk excess emission, have determined IDFs of $\sim$
80--85\% for the Trapezium (Lada et al.  2000) and NGC 2024 (Haisch, Lada \& Lada 2000;
Haisch et al.  2001c) clusters in Orion, two of the youngest clusters studied.  These
results suggest that circumstellar disks are a natural byproduct of the star formation
process and that most stars, independent of mass, are therefore born with the ability
to form planetary systems.

However, infrared studies of embedded populations and clusters have also suggested that
the duration of the accretion or protoplanetary disk phase may be relatively brief (
3-15 $\times$ 10$^6$ yrs; Strom et al.  1989; Skrutskie et al.  1990; Strom, Edwards
\& Skrutskie 1993; Lada \& Lada 1995; Brandner et al.  2000).  Since the building times
for giant gaseous planets are estimated to be 10$^7$ yrs  or longer (Lissauer 2001),
it is critically important to accurately constrain the empirical disk lifetime
measurements.  Young clusters offer an excellent laboratory for investigating disk
evolution timescales.  These clusters present statistically significant samples of
stars whose mean ages are well determined.  Moreover, by combining observations of
embedded as well as revealed clusters one can create a sample that spans a range in age
much greater than that characterizes any individual star forming region.  Recently,
Haisch, Lada \& Lada (2001a) performed the first systematic and homogeneous
observational survey for circumstellar disks in young clusters.  They used JHKL imaging
observations to determine the CDF in six clusters whose ages varied between 0.3-30
Myrs.  Figure~\ref{diskfreq} shows the CDF of their sample along with CDF measurements
for two embedded T associations (Taurus, Kenyon \& Hartmann 1995; Chamaeleon I, Kenyon
\& Gomez, 2001).  The observations show a rapid decline in the CDF with cluster age.
Half of the disks in a cluster appear to be lost within only 2-3 Myrs and essentially
all the disks are gone in about 5-6 Myrs.  Moreover, observations also indicate that
disk lifetimes are also functions of stellar mass with disks around higher mass stars
evolving more rapidly than disks around low mass stars (Haisch, Lada \& Lada 2001b)
Such a rapid timescale for disk evolution places stringent constraints on the timescale
for building planets, particularly giant gaseous planets.  

However, the JHKL observations of Haisch et al (2001a) trace the infrared excess arising from
small (micron-sized), hot (900 K) dust grains located in the inner regions of a circumstellar
disk ($\sim$ 0.25 AU).  It is possible for substantial amounts of material to be still
present in the disk if either this material is purely gaseous or if the disk has a large
inner hole.  The former situation could arise if significant grain growth occurs due dust
settling and rapid grain coagulation and growth, as is the expected first step in the
formation of planetesimals.  However, any turbulence in the disk would be expected to keep
detectable amounts of small dust grains suspended within the gas (Ruden 1999), so dust should
still remain a good tracer of total disk mass as the disks evolve.  Indeed, recent
observations of dust and H$_2$ emission in more evolved and much less massive debris disks
appears to support this assumption (Thi et al.  2001) The latter possibility could occur if
protoplanetary disks evolved large inner holes as they aged.  It has been noted that in the
Taurus clouds, near-infrared L-band excess is closely correlated with millimeter-continuum
emission which traces outer disk material suggesting that the evolution of the inner and
outer disks is homologous and occurs on the same timescale (Haisch et al.  2001b).
Similarly, a recent millimeter continuum survey of 4 young embedded clusters (NGC 2071, NGC
2068, NGC 1333, and IC 348) has found that the variation in the fraction of detected
millimeter sources from cluster to cluster is similar to the variation in the fraction of
near-IR excess sources (Lada, Haisch and Beckwith 2003).  This implies that the lifetimes of
the inner disk, as detected by near-IR excess, and the outer disk, as detected at millimeter
wavelengths, are coupled for embedded clusters as they were for the Taurus population.  These
observations strengthen the previous L band excess determination of a short disk lifetime of
$<$ 6 million years, and further suggest that the lifetime for massive outer disks may be
only as long as 3 million years.  Indeed, the dearth of strong millimeter continuum emission
from the disk population of the very young Trapezium cluster implies even more rapid outer
disk evolution (Mundy, Looney \& Lada 1995; Bally et al.  1998; Lada 1999).  It may also be
possible that disk lifetimes depend on internal cluster environment (e.g., the degree to
which a cluster produces O stars).  Systematic surveys of a larger population of clusters are
needed to assess this possibility.  In particular, mid-infrared surveys that will be carried
out by the SIRTF mission, should more definitively address this question.

\subsection{Brown Dwarfs}

As discussed earlier in this review, embedded clusters contain significant populations of
substellar objects.  Indeed, the Trapezium cluster alone contains nearly as many brown
dwarfs as have been identified in the galactic field to this point in time.  However, in
the two clusters (Trapezium and IC 348) with the most complete substellar census,
substellar objects account for only about 20-25\% of the total cluster population. Thus,
if embedded clusters such as these, supplied the galactic field population, then we 
would expect to find only 1 brown dwarf for every 3-4 stars in a typical volume of
Galactic space. 

Embedded cluster research has provided one of the most interesting discoveries concerning
the nature of brown dwarfs.  Examination of young brown dwarfs and brown dwarf candidates
in a number of embedded clusters, including IC 348, Ophiuchus and the Trapezium and in
other star forming regions, such as the Chamaeleon Clouds, has produced evidence that some
brown dwarfs emit excess emission at near- and mid-infrared wavelengths similar to that
emitted by stars with circumstellar disks (e.g., Wilking et al.  1999, Luhman 1999; Nata \&
Testi 2001; Natta et al.  2002).  Moreover, deep near-infrared images of the Trapezium have
enabled the first statistically significant determination of the disk frequency for an
embedded {\it sub}stellar population.  Muench et al.  (2001) found a very high fraction (65\%) of
the substellar objects in the cluster to display infrared excess emission at 2 $\mu$m
suggestive of the presence of circumstellar disks.  Moreover, Muench et al (2001)
discovered that about 20\% of the substellar population were optical proplyds on HST archive
images, independently confirming the presence of circumstellar disks around a significant
fraction of the substellar objects in the cluster.  Because observations of 2 $\mu$m
excess undercounts disk bearing stars, the actual fraction of substellar objects with disks
is likely greater than 65\% and thus very similar to the disk fraction (80\%) observed for
the stellar population of the cluster (Lada et al.  2000).  Like stars, brown dwarfs appear
to be formed surrounded by disks and as a result possess the ability to form planetary
systems.  The detection of infrared excess around the faintest sources in the KLF of the
Trapezium cluster also conclusively established the nature of these sources as young
objects and cluster members and thus confirmed their status as bona fide substellar
objects.  Furthermore, the detection of a high disk fraction (similar to that of the
stellar population) coupled with the smooth continuity of the IMF across the hydrogen
burning limit provides strong evidence that freely floating brown dwarfs are a natural
product of the star formation process in embedded clusters.  The formation process for
brown dwarfs is essentially identical to that of stars.


\subsection{Binary Stars}

Since most field stars appear to be binary systems understanding the origin of binary star
systems is of fundamental importance for developing a general theory of star formation.
Clusters are important laboratories for investigating binary formation and evolution.  In
this context it is interesting that the binary fractions of well studied embedded clusters
such as the Trapezium (Prosser et al.  1994; Petr et al.  1998; Simon, Close \& Beck 1999),
IC 348 (Duchene, Bouvier \& Simon 1999) and Rho Ophiuchi (Simon et al.  1995) are found to
be indistinguishable from that of the galactic field.  On the other hand, it is well
established that the binary fraction of the embedded population in the Taurus-Auriga
association is significantly (a factor of 2) in excess of that of the galactic field (e.g.,
Leinert et al.  1993; Ghez, Neugebauer, \& Matthews 1993; Duchene 1999).  This difference
in binary fraction provides an important clue relating to the origins of galactic field
stars.  Specifically, it supports the notion that most field stars originated in embedded
clusters rather than in embedded associations such as the Taurus-Auriga clouds.

However, it has been suggested that the Taurus-Auriga binary fraction may represent the
initial binary fraction even for stars that form in embedded clusters.  Kroupa (1995a,b)
performed N-body experiments to simulate the evolution of a binary population in a young
cluster.  These experiments began with clusters containing 100\% binaries and showed that
stellar encounters could disrupt binaries and reduce the overall binary fraction with time.
Moreover, Kroupa found that the field star binary population could be produced by such a
model, if most stars formed in what he identified as a dominant-mode cluster, a cluster
with roughly 200 systems and a half mass radius of 0.8 pc.  The fact that such systems are
relatively common (e.g., Table 1) lends support to this notion.  Furthermore, the concept
that disruption of multiple systems can occur in clusters appears to be supported by the
observation of a deficit of wide binaries in the Trapezium cluster (Scally, Clarke \&
McCaughrean 1999).  However, the binary fraction of embedded clusters is not significantly
different from that of much older open clusters (Patience \& Duchene 2001).  This indicates
that any evolution of the binary fraction must have occurred on timescales of order 1 Myr or
less.  Such rapid evolution in the binary population can occur for a very dense embedded
cluster such as the Trapezium (Kroupa, Petr \& McCaughrean 1999, Kroupa Aarseth, \& Hurley
2001).  It is not clear, however, whether binaries can be as efficiently disrupted in the
lower density clusters which account for most star formation.  The question of a universal
initial binary fraction remains open.  Star formation may produce a variety of outcomes for
the emerging binary fraction due to a corresponding variety of initial conditions.
Determinations of the binary properties of the protostellar populations in embedded
clusters would provide an important test of this question.  If there is a
universal initial binary fraction then essentially all protostellar objects must be nascent
binary systems.  High resolution infrared imaging and spectroscopic monitoring of Class I
sources in embedded clusters could resolve this issue.

\section{ORIGIN AND DYNAMICAL EVOLUTION}

\subsection{Formation of Embedded Clusters}

To understand how an embedded cluster forms we must understand two basic physical
processes:  1) the formation of a massive, dense core in a GMC and 2) the
subsequent development of stars from dense gas in the core.  Molecular clouds form from the
turbulent, diffuse and atomic interstellar medium by a physical process or collection of
processes that are far from understood.  Overall this process likely involves the complex
interplay of such things as spiral density waves, supernova explosions, the galactic
dynamo, phase transitions, and various types of instabilities (e.g, thermal, gravitational,
magneto-hydrodynamic, etc.)  (e.g., Elmegreen 1991, 1993).  The vast majority of GMCs are
observed to contain dense gas and signposts of star formation suggesting that the formation
of dense cores and then stars proceeds very rapidly after the cloud has formed from
the diffuse interstellar medium.  The GMCs that form from the ISM are gravitationally bound
entities with highly supersonic and turbulent velocity fields.  The turbulent dissipation
timescales for GMCs are thought to be shorter than the cloud lifetime suggesting that on
global scales the clouds are stabilized against collapse by internal turbulent pressure.
Numerical simulations (e.g., Klessen, Heitsch \& Mac Low 2000) suggest that under such
conditions supersonic turbulent flows can collide, shock and dissipate energy.  Under the
right conditions these collisions can produce dense cores which are gravitationally
unstable and decouple from the overall turbulent flow.  The largest and most massive of
these fragments are then the potential sites of cluster formation.

The second step of the cluster formation process, the rapid evolution of dense gas in a
massive core to form stars, likely involves the continued dissipation of turbulence in the
dense gas which is followed by, fragmentation, gravitational instability and the formation
of protostellar seeds which grow by accreting their infalling envelopes and then perhaps
other surrounding dense gas from the general potential in which they are embedded (e.g.,
see reviews by Clarke et al.  2001, Elmegreen et al.  2001).  This scenario appears to be
quite different than that which has successfully explained the formation of isolated low
mass stars from individual low mass cores.  Such solitary stars form from initially
turbulent, magnetically supported, dense cores which evolve through ambipolar diffusion of
magnetic fields to be dynamically unstable and then collapse from the inside out (Shu,
Adams \& Lizano 1987).  The cores which form isolated low mass stars in this manner have
sizes that are considerably larger than the separation of stars in an embedded cluster.
Evidently protostellar cores in clusters must have smaller radii than those which form in
isolation.  This suggests that cluster forming cores must experience significant
fragmentation in their evolution to form stars.  The physical mechanism that produces this
fragmentation is not well understood.  This process likely involves progressive cooling of
a marginally stable or collapsing massive core which continuously reduces the Jean's mass.
In turbulent dense cores this cooling takes place as a result of dissipation or loss of
turbulence.  An elegant possibility to account for fragmentation was proposed by Myers
(1998) for the case of MHD turbulence.  If the ionization rate in a massive core is low
enough (i.e., the extinction is high enough that cosmic rays are the sole source of
ionization) then MHD waves greater than a certain frequency cannot couple well to the
neutral gas.  This corresponds to a cutoff wavelength, below which turbulence can no longer
be sustained (e.g., Mouschovias 1991).  This situation can lead to the formation of a
matrix of critically stable Bonnor-Ebert condensations or kernels confined by the pressure
in the surrounding gas.  Myers (1998) finds that for typical conditions the sizes of these
kernels can be comparable to the separation of stars in embedded clusters.  Fragmentation
can also be produced in the turbulent decay process as flows collide and shock, creating
density enhancements, which if massive enough can become gravitationally bound and separate
from the general turbulent velocity field (e.g., Klessen \& Burkert 2000, 2001).

Once these fragments or kernels become gravitationally unstable they collapse, and gaining
mass through infall of surrounding material become protostars.  However, the rates at which
protostellar condensations grow must vary significantly within the cluster.  This is
because the star formation process must produce a range of stellar and substellar masses
spanning three orders of magnitude within a timescale of only a few (1-2) million years in
order to reproduce the stellar IMF.  As they move through the cluster core protostellar
fragments also accrete additional material from the reservoir of residual gas not locked up
in other protostellar objects (Bonnell et al.  2001a).  Since all these stellar embryos
share a common envelope, a process of competitive accretion begins with initially more
massive protostellar clumps or clumps closer to the center of the cluster experiencing
higher accretion rates.  The process is highly nonlinear and, even for a cluster with
initially equal mass protostellar fragments, can lead to the development of a protostellar
mass spectrum similar to that of the stellar IMF (Bonnell et al.  2001b; Klessen 2001).  In
this picture the more massive stars tend to be formed in the central regions of the cluster
leading to some degree of primordial mass segregation.  It is also possible for
protostellar fragments in the dense inner regions of the cluster to collide and coalesce
leading to the production of very massive stars (Bonnell, Bate \& Zinnecker 1998).  It
otherwise would be difficult to built up a massive star from general accretion since
radiation pressure from embryonic stars more massive than about 10 \msun can reverse infall
and stunt the growth of the star (e.g., Adams, Lada \& Shu 1987).

As indicated in Table~1 embedded clusters are found to span a range in mass of at least two
orders of magnitude.  Yet their ages are all probably within a factor of 2 of 2 Myr.
This fact suggests that the star formation rates in clusters vary widely.  This is even more
apparent when one considers the more distant "star burst" clusters such NGC 3603 and R 136
which contain almost as many O stars as the entire memberships of embedded clusters like
NGC 1333 and Ophiuchi. Yet all these clusters are likely within a factor of 2 of the same
age.  The finding of a significantly higher star formation rate for the Trapezium cluster
compared to that in the two embedded clusters (IC 348 \& NGC 1333) in the Perseus molecular
complex led Lada et al. (1996) to suggest that the star formation in the Trapezium
cluster was externally triggered.  Compressive triggering increases the external pressure
and thus the density of a core and in doing so speeds up the star formation.  The close
association of other clusters with adjacent HII regions has also suggested that triggering
may have played an important role in the formation of at least some clusters (e.g., S 255,
Howard et al.  1997; S 281, Megeath \& Wilson 1997, W3-W4, Carpenter et al.  2000).
At some point, when molecular gas is either depleted or expelled from the cluster, 
star formation ceases and the cluster emerges from its molecular womb.

\subsection{Emergence from Molecular Clouds: Dynamical Evolution \& Infant Mortality}

Although the origin of embedded clusters remains a mystery, the subsequent dyanamical
evolution of embedded clusters and their emergence from molecular clouds are well posed
theoretical problems which have been studied extensively both analytically (e.g., Hills
1980; Elmegreen 1983; Verschueren \& David 1989) and numerically (Lada, Margulis \&
Dearborn 1984; Goodwin 1997; Geyer \& Burkert 2001; Kroupa, Aarseth \& Huley 2001; Kroupa
\& Boily 2002).  As described earlier, clusters form in massive, dense cores of molecular
gas which are strongly self-gravitating.  Star formation is an inherently destructive
process and upon formation, new stars will immediately begin to disrupt their surrounding
gaseous environments.  The birth of high mass stars can be particularly destructive and not
only lead to the rapid disruption of a cluster forming core but in addition to the complete
dispersal of an entire GMC (e.g, Whitworth 1979).  Moreover, outflows generated by a
population of low mass stars are also capable of disrupting a massive cloud core in a
relatively short time (e.g., Matzner \& McKee 2000).  As a result of these effects, star
formation turns out to be a relatively inefficient process.  Thus, the gravitational glue
that binds together the system of stars and gas in an embedded cluster may be largely
provided by the gas.  Stars are then expected to orbit in a deep potential well of the
dense core with orbital velocities (i.e., $\sigma \approx (G[M_{stars} + M_{gas}]/R)^{0.5}$)
characteristic of the virial velocities of the dense gaseous material.  As it emerges from
a cloud the evolution of an embedded stellar cluster is consequently sensitively coupled
to the evolution of its surrounding gas.

The two physical parameters that determine the evolution of an emerging embedded cluster
are the star formation efficiency and the timescale of gas dispersal from the cluster.  The
star formation efficiency (SFE $ = M_{stars}/(M_{gas} + M_{stars})$ ) is a fundamental
parameter of both the star and cluster formation processes.  Since the measurement of the SFE
requires a reliable and systematic determination of {\it both} the gaseous and stellar mass
within a core, accurate SFE measurements are not generally available for cluster forming
regions.  In Table 2 we list the star formation efficiencies for the sample of
nearby embedded clusters (drawn from Table 1) which appear to be fully embedded
and for which reasonable empirical determinations for the gaseous and stellar mass exist.
The SFEs range from approximately 10-30\%.  These estimates are typical of those sometimes
estimated for other embedded clusters but are also significantly higher than the global
SFEs estimated for entire GMCs which are typically only 1-5 \% (e.g., Duerr, Imhoff \& Lada
1982).  The fact that the clusters with the lowest SFEs (Serpens, Ophiuchi, NGC 1333)
appear to be the least evolved, suggests that the SFE of a cluster increases with time and
can reach a maximum value of typically 30\% by the time the cluster emerges from its
parental cloud core.  Whether all clusters can reach SFEs as high as 30\% before emerging
from a molecular cloud is unclear, however, it does seem apparent that clusters rarely
achieve SFEs much in excess of 30\% before emerging from molecular clouds.


\begin{table}[t]
\centerline{Table 2: Star Formation Efficiencies for Nearby Embedded Clusters}  
\label{sfe}
\begin{center}
\begin{tabular}{@{}lcccc@{}} 
\hline
Cluster Name & Core Mass (\msunp) & Stellar Mass (\msunp) &\ SFE &\ Ref.\\
\hline
Serpens & 300 & \ 27 & \ 0.08 & 1\\ 
Rho Oph & 550 & \ 53 & \ 0.09 & 2\\
NGC 1333 & 950 & \ 79 & \ 0.08 & 3\\
Mon R2 & 1000 &  341 & \ 0.25 & 4\\
NGC 2024 & 430 & 182 &\ 0.33 & 5\\
NGC 2068 & 266 & 113 &\ 0.30 & 5\\
NGC 2071 & 456 &\ 62 &\ 0.12 & 5\\
\hline
\end{tabular} 
\end{center}
References: 1-Olmi \& Testi 2002; 2-Wilking \& Lada 1983; 3-Warin et al. 1996;
4-Wolf, Lada \& Bally 1990; 5-Lada et al 1991b.
\end{table}

The timescale for gas removal, $\tau_{gr}$, from a cluster is even less well constrained by
empirical data than the SFE.  However, the ultimate dynamical fate of an embedded cluster
is determined by the relationship between $\tau_{gr}$ and $\tau_{cross}$, the initial
dynamical timescale of the cluster (i.e., $\tau_{cross} = 2R/\sigma$, where $R$ is the
radius and $\sigma$ the velocity dispersion of the cluster).  There exist two important
dynamical regimes for $\tau_{gr}$ corresponding to explosive ($\tau_{gr} << \tau_{cross}$)
and adiabatic ($\tau_{gr} >> \tau_{cross}$) gas removal times.  Typical embedded clusters
are characterized by $\tau_{cross} \sim$ 1 Myr.  Clusters that form O stars will likely
remove any residual gas on a timescale shorter than this dynamical time.  O stars can
quickly ionize and heat surrounding gas to temperatures of 10$^4$~K causing an abrupt
increase in pressure which results in rapid expansion of the gas.  The expansion velocities
are on the order of the sound speed in the hot gas and for the dimensions of embedded
clusters correspond to gas removal timescales that can be as short as a few times 10$^4$
years.  The dynamical response of the stars which are left behind after such explosive gas
removal will depend on the SFE achieved by the core at the moment of gas dispersal.  The
condition for the cluster to remain bound in the face of rapid gas removal is that the
escape speed from the cluster, $V_{esc} \approx (2GM/R)^{0.5}$, is less than $\sigma$, the
instantaneous velocity dispersion of the embedded stars at the time of gas dispersal.  Thus
a bound group will emerge only if the SFE is greater than 50\% (Wilking \& Lada 1983).
Consequently, the fact that the SFEs of embedded clusters are always observed to be less
than 50\% is critically important for understanding their dynamical evolution.  Apparently,
it is very difficult for embedded clusters to evolve to bound open clusters, particularly
if they form with O stars.

However, classical open clusters, like the Pleaides, do exist in sufficient numbers that at
least some embedded clusters with SFEs less than 50\% must have evolved to become
relatively long-lived, bound systems.  For slow gas removal times, $\tau_{gr} >
\tau_{cross}$, clusters even with low SFEs can adiabatically adjust and expand to new
states of virial equilibrium and remain bound.  The fact that clusters older than about 5
Myr are observed to rarely be associated with molecular gas suggests that $\tau_{gr} <$ 5
Myr.  This is close enough to the crossing timescales that numerical calculations are
necessary to investigate the response of clusters to this slow gas removal.  Moreover, to
produce a bound group which is stable against galactic tides and the tidal forces of its
parental GMC requires additional stringent constraints on the initial conditions prevailing
in the cluster forming cloud core and on $\tau_{gr}$ (Lada et al.  1984).  To evolve to a
bound cluster like the Pleiades, the typical embedded cluster would have to have a gas
removal time of at least a few (3-4) crossing times, which corresponds to a few million
years for typical conditions.  This timescale is of the same order as the cluster formation
or gestation time and would require a cluster to be losing mass while simultaneously
forming stars.  Outflows generated by low mass stars can remove gas during the star
formation process and Matzner \& McKee (2000) have shown that such outflows can completely
disrupt cluster forming cores for SFEs in the 30-50\% range.  Whether such outflow driven
mass dispersal can occur over as long a timescale as required is not clear.  This depends
on the detailed star formation history of the cluster, in particular the star formation
rate and its variation in time and neither of these quantities is sufficiently well
constrainted by existing observations.

Numerical calculations show that those embedded clusters which do evolve to bound systems
in such a manner undergo significant expansion as they emerge from a cloud and consequently
one expects bound open clusters to have significantly larger radii than embedded clusters,
as is observed.  In addition, during emergence clusters can expand for long periods before
they reach a final equilibrium.  The appearance of bound and unbound emerging clusters are
indistinguishable for clusters with ages less than 10 Myrs.  Numerical calculations also
show that even clusters that can survive emergence from a molecular cloud as  bound
systems may lose 10\% to 80\% of their numbers in the process.  The more violent the gas
disruption the smaller the fraction of stars that are bound.  However, even clusters that
experience explosive loss of gas can leave behind bound cores containing 10-20\% of the
original stellar population (Lada et al 1984; Kroupa \& Boily 2002).  Therefore, bound
clusters, even with O stars, can be produced provided that the progenitor embedded cluster was
substantially more massive and dense than the surviving open cluster.

Kroupa \& Boily (2002) have pointed out that classical open clusters such as the Pleiades
and Praesepe are massive enough to have formed with O stars and have posited that
Pleiades-like clusters formed from much more massive protoclusters, most of whose original
members were lost in the emergence from molecular gas.  Indeed, Kroupa Aarseth \& Hurley
(2001) have identified the Trapezium-ONC cluster as such a possible proto-Pleiades system
with the Trapezium cluster likely being the future bound remnant of the emerging, mostly
unbound ONC cluster.  However, such proto-Pleiades systems would need to initially contain
$\sim$ 10$^4$ stars, considerably more than appear to be actually contained in the ONC
(1700 stars) and RCW 38 (1300 stars), the richest clusters in Table 1.  In
addition, if most open clusters lost half or more of their original stars upon emergence
from a molecular cloud, it would be difficult to understand both the similarity of the
embedded cluster and open cluster mass functions as well as the similarity of the stellar
IMFs of embedded and open clusters.  On the other hand, a number of very rich and massive
(10$^{4-5}$ \msunp) clusters in the Large Magellanic Cloud have been found to be surrounded
by halos of unbound stars which account for as much as 50\% of their total masses (Elson et
al.  1987).  If galactic globular clusters formed in such a manner, the stars lost in their
emergence from their parental clouds could account for all the Population II field stars
in the Galactic halo (Kroupa \& Boily 2002).  The numerical simulations appear to indicate
that under most conditions leading to the formation of bound clusters, a significant
fraction of the initial stellar population of a protocluster will be lost.  Clearly more
extensive observations of the near environments of emerging embedded clusters would be
useful to better ascertain and quantify the extent of any distributed or extended halos of
young stars around these objects.  Such information would provide important constraints for
future modeling.

The production of a bound cluster from a dense cloud core clearly requires very special
physical conditions.  This must be a rare occurrence.  The observed low SFEs for embedded
clusters can account for the high infant mortality rate of clusters inferred from the
relatively large numbers and high birthrates of embedded clusters compared to classical
open clusters.  Most ($\sim$ 90-95\%) embedded clusters must emerge from molecular clouds
as unbound systems.  Only the most massive ($M_{EC} \geq $ 500 \msunp) embedded clusters
survive emergence from molecular clouds to become open clusters.  Thus, although
most stars form in embedded clusters, these stellar systems evolve to become the members of
unbound associations, not bound clusters.  However, bound classical clusters form at a
sufficiently high rate that on average, each OB association (and GMC complex) probably
produces one such system (Elmegreen \& Clemens 1985) accounting for about 10\% of all stars
formed within the Galaxy (Roberts 1957; Adams \& Myers 2001).


\section{CONCLUDING REMARKS}

The discovery of large numbers of embedded clusters in molecular clouds over the last
fifteen years has lead to the realization that these young protoclusters are responsible
for a significant fraction of all star formation currently occurring in the Galaxy.
Embedded clusters  may
very well be the fundamental units of star formation in GMCs.  Conceived in the mysterious
physical process that transforms diffuse interstellar matter into massive and dense
molecular cloud cores, embedded clusters are born at a rate that significantly exceeds that
estimated for classical open clusters.  Evidently, the vast majority of embedded clusters
do not survive their emergence from molecular clouds as bound stellar systems.  Their high
infant mortality rate is mostly the result of the low to modest star formation efficiency
and rapid gas dispersal which characterizes their birth.  There are more than 20 embedded
clusters formed for every cluster born that ultimately evolves into a long-lived system
like the Pleiades.  As the primary sites of star birth in molecular clouds,
embedded clusters are important laboratories for studying the origin and early
evolution of stars and planetary systems.  The fundamental properties of the Galactic
stellar population, such as its IMF, its stellar multiplicity, 
and the frequency of planetary systems within it, are forged in embedded clusters.

Although, observations over the last fifteen years have clearly established the the central
role of embedded clusters in the star formation process, the fundamental parameters of
these extremely young clusters are still very poorly constrained.  In particular, the
overall census of embedded clusters is far from complete, even within 1 Kpc of the Sun.  In
addition, very little information concerning the ages of embedded clusters exists.
Accurate information concerning the spatial sizes, number of members, masses, and distances
for most embedded clusters is also lacking.  Nonetheless, despite these deficiencies,
studies of individual embedded clusters have provided new insights into fundamental
astrophysical problems, such as determining the functional form and universality of the
IMF, the frequency and lifetimes of protoplanetary disks, and the ubiquity and nature of
brown dwarfs.  However, to date only a small number of such clusters have been studied in
any detail.  It remains to be determined whether the trends determined for this small
sample are representative of the majority of embedded clusters and star formation events in
the Galaxy as a whole.  For example, does the IMF of the Trapezium cluster truly represent
a universal IMF?  Is the fraction of freely floating brown dwarfs always approximately
20-25\% of a cluster population?  Is the circumstellar disk lifetime the same in all
clusters?  Other important questions also remain open.  Do the progenitors of bound open
clusters ever contain O stars?  How frequently do O stars form in embedded clusters?  Is
the primordial binary fraction the same for stars formed in and outside rich clusters?
What is the most massive embedded cluster that can be formed from a GMC?  How many such
clusters exist in the Galaxy?  What is the actual number of poor (N$_*$$<$ 35) embedded
clusters or stellar aggregates formed and what is the fraction of all stars produced in
such groups?  Is the process that produces embedded clusters in any way related to that
responsible for the formation of globular clusters?

Resolving these issues will require an extensive effort in both observation and theory.
Prospects for progress continue to be bright due to the development of important new
observational capabilities.  These include, wide field infrared imaging and multi-object
spectroscopy using large, ground-based telescopes, airborne and space-based infrared
imaging and spectroscopy provided by missions such as SOFIA, SIRTF and NGST, and NIR all
sky surveys such as 2MASS and DENIS.  Practically everything we know about embedded
clusters we have learned in the last fifteen years.  It is difficult to predict but
exciting to contemplate what will be learned in the next fifteen years as a result of these
new capabilities and the continued dedicated efforts of astronomers who work on these
problems.  However, whatever the outcome of such research, there can be little doubt that
the result of these efforts will be to enrich our understanding of the star and planet
formation process in the universe.





%
%
%

\noindent
ACKNOWLEDGEMENTS

We are grateful to Jo\~ao Alves, Richard Elston and August Muench for assistance in
preparation of figures.  We thank also August Muench and Richard Elston for many useful
comments and criticisms of earlier versions of this manuscript.  EAL acknowledges support
from a Presidential Early Career Award for Scientists and Engineers (NSF AST 9733367) to
the University of Florida.  Finally, we gratefully acknowledge the efforts of the 
many researchers whose contributions have made this such an interesting and stimulating
subject to review. 







\begin{table}[t]
\centerline{Table 1: Catalog of Embedded Clusters}
\begin{center}
\begin{tabular}{llccrrrcrl}
\hline  
\hline  
EC & Name &   RA         &     Dec         &    Distance   &     Size  &      N$_*$ &    K  &     Mass & Ref \\
     &   (J2000)    &      (J2000)    &     (pc)      &      (pc) &            &    (limit)          &    M$_\odot$ &       \\
\hline  
1   & NGC281W       & 00:52:23.7   &  +56:33:45 &   2100 &         & 231  &  18.0    &    130   & 1,2 \\
2   & NGC 281E      & 00:54:14.7   &  +56:33:22 &   2100 &         & 88   &  17.0    &    57    & 1 \\
3   & 01546+6319    & 01 58:19.8   &  +63:33:59 &   2400 &   0.54  & 54   &  17.5  &    35    & 3 \\
4   & 02044+6031    & 02:08:04.7   &  +60:46:02 &   2400 &   0.73  & 147  &  17.5  &    94    & 3 \\
5   & 02048+5957    & 02:08:27.0   &  +60:11:46 &   2400 &   0.56  & 58   &  17.5  &    37    & 3 \\
6   & 02054+6011    & 02:09:01.3   &  +60:25:16 &   2400 &   0.59  & 70   &  17.5  &    45    & 3 \\
7   & 02175+5845    & 02:21:07.7   &  +58:59:06 &   2400 &   0.73  & 109  &  17.5  &    70    &  \\
8   & IC 1805W      & 02:25:14.5   &  +61:27:00 &   2300 &         & 79   &  17.0    &    57    & 1 \\
9   & 02232+6138    & 02:27:04.1   &  +61:52:22 &   2400 &   0.91  & 205  &  17.5  &    130   & 3 \\
10  & 02245+6115    & 02:28:21.5   &  +61:28:29 &   2400 &   0.64  & 121  &  17.5  &    77    & 3 \\
11  & 02407+6047    & 02:44:37.8   &  +60:59:53 &   2400 &   0.46  & 50   &  17.5  &    32    & 3 \\
12  & 02461+6147    & 02:50:09.2   &  +61:59:58 &   2400 &   0.72  & 115  &  17.5  &    73    & 3 \\
13  & 02484+6022    & 02:52:18.7   &  +60:34:59 &   2400 &   0.62  & 86   &  17.5  &    55    & 3 \\
14  & 02497+6217    & 02:53:43.2   &  +62:29:23 &   2400 &   0.38  & 36   &  17.5  &    23    & 3 \\
15  & AFGL 4029     & 03:01:32.3   &  +60:29:12 &   2200 &         & 173  &  16.5  &    140   & 1 \\
16  & 02541+6208    & 02:58:13.2   &  +62:20:29 &   2400 &   0.45  & 40   &  17.5  &    26    & 3 \\
17  & W3IRS5        & 02:25:40.6   &  +62:05:52 &   2400 &         & 87   &  17.0    &    64    & 4 \\
18  & 02570+6028    & 03:01:00.7   &  +60:40:20 &   2400 &   0.62  & 78   &  17.5  &    50    & 3 \\
19  & 02575+6017    & 03:01:29.2   &  +60:29:12 &   2400 &   1     & 240  &  17.5  &    150   & 3 \\
20  & 02593+6016    & 03:03:17.9   &  +60:27:52 &   2400 &   0.62  & 88   &  17.5  &    56    & 3 \\
21  & AFGL 437      & 03:07:25.6   &  +58:30:52 &   2000 &         & 122  &  17.0    &    79    & 1 \\
22  & AFGL 490      & 03:27:38.7   &  +58:46:58 &   900  &         & 45   &  16.5  &    25    & 1 \\
23  & NGC 1333      & 03:32:08.1   &  +31:31:03 &   318  &   0.49  & 143  &  14.5  &    79    & 5 \\
24  & IC 348 	      & 03:44:21.5   &  +32:10:16 &   320  &         & 300  &  15.0    &    160   & 6,7  \\
25  & LKHalpha 101  & 04:30:14.4   &  +35:16:25 &   800  &         & 150  &  15.0    &    98    & 8,9 \\
26  & S242          & 05:52:12.9   &  +26:59:33 &   2100 &         & 96   &  16.5  &    81    & 1 \\

\hline  
\end{tabular}
\end{center}
\end{table}
\vfill\eject

\begin{table}
\centerline{Table 1: Catalog of Embedded Clusters (Continued)}
\begin{center}
\begin{tabular}{llccrrrcrl}
\hline  
\hline  

EC & Name &   RA         &     Dec         &    Distance   &     Size  &      N$_*$ &    K  &     Mass & Ref \\
     &   (J2000)    &      (J2000)    &     (pc)      &      (pc) &            &    (limit)          &    M$_\odot$ &       \\
\hline  
27  & AFGL 5142     &  05:30:45.6  &   +33:47:51 &    1800 &       &  60 &    16.0   &     50  &   1,2  \\
28  & OMC 2  	&	05:35:27.3   &  --\ 05:09:39  &   500  &         & 119  &  17.5  &    63    & 1 \\
29  & L1641 N	&	05:36:23.1   &  --\ 06:23:40  &   500  &   0.66  & 43   &  14.7 &    24    & 1,10  \\
30  & ONC/Trapezium  & 05:37:47.4  &   --\ 05:21:46 &    500 &    3.8  &  1740&   14.0   &    1100 & 11,12  \\
    & Trapezium      & 05:37:47.4  &   --\ 05:21:46 &    500 &   0.47    &  780 &   17.5 &    413  & 13 \\
31  & AFGL 5157      & 05:37:47.8  &   +31:59:24&    1800&         &  71  &   16.5 &    50   & 1 \\
32  & L1641C  &	05:38:46.9   &  --\ 07:01:40 &   500  &   0.66  & 47   &  14.7 &   27    & 1,10 \\
33  & S235B          & 05:40:52.5  &   +35:41:25   &  1800 &      &   300  &  16.5    &     220 &  1,2 \\
34  & NGC 2024       & 05:41:42.6  &   --\ 00:53:46  &  400 &    0.88 &  309 &   14.0   &   180  &  14  \\
35  & NGC 2068       & 05:46:41.8  &   +00:06:21  &  400 &    0.86 &  192 &   14.0   &   110  &  14  \\
36  & NGC 2071       & 05:47:10.0  &   +00:19:19  &  400 &    0.59 &  105 &   14.0   &   60   &  14  \\
37  & L1641S 	&	05:52:28.9   &  --\ 08:07:30   & 500  &   1.3   & 134  &  14.7 &   78    & 1,10  \\
38  & MWC 137	&	06:18:45.5   &  +15:16 52   & 1300 &   0.4   & 59   &  16.5  &   35    & 15  \\
39  & NGC 2282       & 06:46:51.0  &   +01:18 54  &  1700&    1.6  &  111 &   15.0   &  170  &  16 \\
40  & MonR2  	&	06:07:46.6   &  --\ 06:22:59   &  800 &    1.85 &  371 &   14.0   &     340  &  1,17  \\
41  & AFGL 6366S   &  06:08:40.9  &   +21:31:00  &   1500 &      &   550  &  18.0    &   300 &  1  \\
42  & Gem4    &	06:08:41.0   &  +21:30:49    & 1500 &   1.74 &  114 &   14.5 &   190  &  18  \\
43  & AFGL 5180      & 06:08:54.1  &   +21:38:24   & 1500 &        &   94 &    16.5 &   60  &  1  \\
44  & Gem1    &	06:09:05.4   &  +21:50:20    & 1500 &   1.22 &  56  &   14.5 &   95   &  18 \\
45  & IRAS 06068+2030 & 06:09:51.7 &    +20:30:04  & 1500 &   0.39 &  59  &   15.3 &   54   &  2  \\
46  & GGD 12-15      & 06:10:50.9  &   --\ 06:11:54   &  800 &   1.13 &  134 &   16.5 &   73   & 1,11 \\
47  & IRAS 06155+2319& 06:18:35.1  &   +23:18:11   &  1600&   0.43 &  38  &   15.3 &   41   &  2 \\
48  & RNO 73 	&	06:33:31.0   &  +04:00:06    & 1600 &        & 43   &  16.5  &   28    & 1 \\
49  & NGC 2244      &  06:34:55.3  &   +04:25:13   &  1600&        &  150+&       &      &  19 \\
50  & NGC 2264      &  06:41:03.2  &   +09:53:07   &  800 &        &  360 &   14.0  &    330  &  1,20 \\
51  & S287 N 	&	06:47:50.4   &  --\ 02:12:54    & 1400 &        & 46   &  16.5 &    29    & 1 \\
52  & BSF 56 	&	06:59:14.4   &  --\ 03:54:51    & 1400 &        & 64   &  16.5 &    40    & 1 \\

\hline  
\end{tabular}
\end{center}
\end{table}
\vfill\eject

\begin{table}
\centerline{Table 1: Catalog of Embedded Clusters (Continued)}
\begin{center}
\begin{tabular}{llccrrrcrl}
\hline  
\hline  

EC & Name &   RA         &     Dec         &    Distance   &     Size  &      N$_*$ &    K  &     Mass & Ref \\
     &   (J2000)    &      (J2000)    &     (pc)      &      (pc) &            &    (limit)          &    M$_\odot$ &       \\
\hline  

53  & S 287 C &	06:59:36.6   &  --\ 04:40:22    & 1400 &        & 50   &  16.5 &    31    & 1  \\
54  & L 1654  &	06:59:41.7   &  --\ 07:46:29    & 1100 &        & 415  &  17.0   &    230   & 1 \\
55  & BIP 14  &	08:15:14.8   &  --\ 04:04:41    & 1400 &        & 98   &  16.5 &    61    & 1 \\
56  & RCW38   &	08:59:05.4   &  --\ 47:30:42    & 1700 &        & 1300 &  18.0   &    730   & 1 \\
57  & NGC 3576       & 11:11:57.0  &   --\ 61:18:54   &  2400&        &  51  &   13.0  &    720  & 22 \\
58  & Rho Oph &	16:27:01.6   &  --\ 24:36:41    & 125  &        & 100  &  14.0   &    53    & 23 \\
58  & NGC 6334I      & 17:20:53.0  &   --\ 35:46:57   &  1700&   0.6  &  93  &   16.0  &    78   & 24 \\
60  & Trifid/ M20    & 18:02 23.0  &   --\ 23:01:48   &  1600&        &  85  &   14.3&    190  &  25 \\
61  & M16/NGC6611    & 18:18 48.0  &   --\ 13:47:00   &  1800&        &  300 &   14.0  &    960  &  26 \\
62  & M17     &	18:20 26.0   &  --\ 16:10:36    & 1800 &        & 100  &  12.8 &    890   & 27  \\
63  & NGC 6530/M8    & 18:07:51.9  &   --\ 24:19:32   &  1800&        &  100+&       &          &  28 \\
64  & MWC 297 &	18:27:39.5   &  --\ 03:49:52    & 450  &  0.5   & 37   &  16.7 &    20    & 15 \\
65  & Serpens SVS2   & 18:29:56.8  &   +01:14:46   &  250 &        &  51  &   15.5&    27   &  1,29 \\
66  & S87E           & 19:46:19.9  &   +24:35:24   &  2110&        &  101 &   15.2&    180  &  30 \\
67  & R CrA   &	19:01:53.9   &  --\ 36:57:09    & 700  &        & 40   &  16.5 &    22    & 1,31 \\
68  & S88B    &	19:46:47.0   &  +25:12:43    & 2000 &        & 98   &  15.5 &    120   & 1 \\
69  & S 106  	 &	20:27:25.0   &  +37:21:40    & 600  &  0.3   & 160  &  14.0   &    120   & 32 \\
70  & W 75 N  &	20:38:37.4   &  +42:37:56    & 2000 &        & 130  &  16.5 &    99    & 1 \\
71  & L1228   &	20:57:13.0   &  +77:35:46    & 150  &        & 47   &  18.0   &    25    & 1 \\
72  & IC 5146 &	21:02:36.3   &  +47:27:59    & 1200 &        & 100  &       &          & 33 \\
73  & L988 e  &	21:03:57.5   &  +50:14:38    & 700  &        & 46   &  14.5 &    32  & 1 \\
74  & LKHalpha234    & 21:43:02.2  &   +66:06:29   & 1000 &        & 139  &  17.0   &    76  & 1 \\
75  & Cep A   &	22:56:19.0   &  +62:01:57    & 700  &        & 580  &  17.0   &    310  & 1 \\
76  & Cep C   &	23:05:48.8   &  +62:30:02    & 750  &        & 110  &  16.5 &    60  & 1  \\

\hline  
\end{tabular}
\end{center}
References: 1-Hodapp 1994; 2-Carpenter et al. 1993; 3-Carpenter, Heyer \& Snell 2000; 4-Megeath et al. 1996; 5-Lada, Alves \& Lada 1996;
6-Muench et al. 2003; 7-Lada \& Lada 1995; 8-Aspin \& Barsony 1994; 9-Barsony et al. 1991; 10-Strom, Strom \& Merrill 1993; 11-Carpenter 2000;
12-Hillenbrand \& Carpenter 2000; 13-Muench et al. 2002; 14-Lada E. et al. 1991; 15-Testi, Palla \& Natta 1998; 16-Horner, Lada \& Lada 1997;
17-Carpenter et al. 1997; 18-Carpenter, Snell \& Schloerb 1995; 19-Marshal, van Altena \& Chiu 1992; 20-Lada, Young \& Greene 1993;
21-Alves 2002, private comm.; 22-Persi et al. 1994; 23-Kenyon, Lada \& Barsony 1998; 24-Tapia et al. 1996; 25-Rho et al. 2001; 26-Chini, Kruegel \& Wargau 1992;
27-Lada C. et al. 1991; 28- van den Ancker et al. 1997; 29-Eiroa \& Casali 1992; 30-Chen et al. 2002; 31-Wilking et al 1997; 32 Hodapp \& Rayner 1991;
33-Herbig \& Dahm 2002.
\end{table}

\end{document}